\documentclass[fleqn,usenatbib]{mnras}

\usepackage[T1]{fontenc}
\usepackage{ae,aecompl}

\usepackage[normalem]{ulem}
\usepackage{amsmath}
\usepackage{graphicx} 
\usepackage{lscape}
\usepackage{indentfirst}
\usepackage{enumitem}
\usepackage{xspace}
\usepackage{soul}
\usepackage{amssymb}
\usepackage[dvipsnames]{xcolor}
\usepackage{url}
\usepackage[flushleft]{threeparttable}
\usepackage{bm}

\newcommand {\bc}{\begin {center}}
\newcommand {\ec}{\end {center}}
\newcommand {\beq}{\begin {equation}}
\newcommand {\eeq}{\end {equation}}
\newcommand {\beqa}{\begin {eqnarray}}
\newcommand {\eeqa}{\end {eqnarray}}
\newcommand {\comment}[1]{}

\renewcommand{\d}{{\mathrm{d}}}

\newcommand {\ergs}{\mbox{erg s$^{-1}$}}

\title[Neutrino beaming in ULX pulsars]
{
Neutrino beaming in ultraluminous X-ray pulsars as a result of gravitational lensing by neutron stars
}
\author[A.~A.~Mushtukov et al.] 
{A.~A.~Mushtukov,$^{1}$\thanks{E-mail: alexander.mushtukov@physics.ox.ac.uk (AAM)}
A.~Y. Potekhin,$^{2,3}$
I.~D. Markozov,$^{2,3}$ 
S. Nallan,$^4$ 
K. Kornacka,$^5$
\newauthor
I.~S.~Ognev,$^6$
V. Kravtsov,$^7$
A.~A. Dobrynina,$^6$
A.~D. Kaminker$^2$
\\ 
$^1$ Astrophysics, Department of Physics, University of Oxford, Denys Wilkinson Building, Keble Road, Oxford OX1 3RH                        , UK \\ 
$^2$ Ioffe Institute, Politekhnicheskaya 26, St Petersburg 194021, Russia \\
$^3$ Space Research Institute (IKI) of the Russian Academy of Sciences, Profsoyuznaya 84/32, Moscow  117997, Russia \\
$^4$ Carpe Diem Academy, 6712 Tannahill Drive, San Jose, California, USA \\
$^5$ Christa McAuliffe Academy School of Arts and Sciences,  5200 SW Meadows Rd. Ste. 150, Lake Oswego, OR 97035, USA \\
$^6$ P.G. Demidov Yaroslavl State University, Sovietskaya 14, 150003 Yaroslavl, Russia \\
$^7$ Department of Physics and Astronomy,  FI-20014 University of Turku, Finland \\
} 

\pubyear{2024}

\begin{document}
\label{firstpage}
\pagerange{\pageref{firstpage}--\pageref{lastpage}}
\maketitle

%%%%%%%%%%%%%%%%%%%%%%%%%%%%%%%%%%%%%%%%%%%%%%%%%%

\begin{abstract} 
{
X-ray pulsars experiencing extreme mass accretion rates can produce neutrino emission in the MeV energy band. 
Neutrinos in these systems are emitted in close proximity to the stellar surface and subsequently undergo gravitational bending in the space curved by a neutron star. 
This process results in the formation of a distinct beam pattern of neutrino emission and gives rise to the phenomenon of neutrino pulsars.
The energy flux of neutrinos, when averaged over the neutron star's pulsation period, can differ from the isotropic neutrino energy flux, which impacts the detectability of bright pulsars in neutrinos.
We investigate the process of neutrino beam pattern formation, accounting for neutron star transparency to neutrinos and gravitational bending.
Based on simulated neutrino beam patterns, we estimate the potential difference between the actual and apparent neutrino luminosity. 
We show that the apparent luminosity can 
greatly
exceed the actual luminosity, albeit only in a small fraction of cases, depending on the specific equation of state and the mass of the star.
For example, the amplification can exceed a factor of ten for $\approx0.05\%$ of typical neutron stars with mass of $1.4\,M_\odot$.
Strong amplification is less probable for neutron stars of higher mass.
In the case of strange stars, a fraction of high energy neutrinos can be absorbed and the beam pattern, as well as the amplification of apparent neutrino luminosity, 
depend on neutrino energy.
}
\end{abstract}


%%%%%%%%%%%%%%%%%%%%%%%%%%%%%%%%%%%%%%%%%%%%%%%%%
\section{Introduction}
\label{sec:Intro}
%%%%%%%%%%%%%%%%%%%%%%%%%%%%%%%%%%%%%%%%%%%%%%%%%

X-ray pulsars (XRPs) are accreting neutron stars (NSs) in close binary systems (see \citealt{2022arXiv220414185M} for review). 
Typical field strength at the NS surface here is expected to be $\sim 10^{12}$~G or even stronger.
Such a strong magnetic field modifies the geometry of accretion flow directing it towards small regions located close to the poles of a NS and dramatically influences elementary processes of radiation/matter interaction (see \citealt{2006RPPh...69.2631H}).
The luminosity of XRPs is powered by the accretion process, the most efficient mechanism of energy release.
The apparent luminosity of XRPs covers about nine orders of magnitude.
The lowest detected luminosity is known to be $\sim 10^{32}\,\ergs$.
The brightest XRPs show luminosity $\gtrsim 10^{40}\,\ergs$ and belong to the class of ultra-luminous X-ray sources (see, \citealt{2014Natur.514..202B,2017Sci...355..817I}, and \citealt{2021AstBu..76....6F} for review). 

The geometry of the emission regions at the NS surface is expected to be dependent on the mass accretion rate \citep{1975A&A....42..311B}.
At a relatively low mass accretion rate ($\lesssim 10^{17}$
{g~s}$^{-1}$), 
the accretion flow reaches the stellar surface and is decelerated in the atmosphere of a NS due to the Coulomb collisions, which leads to the formation of hot spots located close to magnetic poles of a star.
At higher mass accretion rates, the luminosity of a NS is sufficiently high ($\gtrsim 10^{37}\,\ergs$) to cause radiative force that stops accretion flow above the stellar surface.
It leads to the formation of accretion columns -- extended structures confined by a strong magnetic field and supported by the radiation pressure gradient \citep{1981A&A....93..255W,2015MNRAS.454.2539M,2022MNRAS.515.4371Z}.
At mass accretion rates exceeding $\sim 10^{19}\,\mathrm{g\,s^{-1}}$, accretion columns can be advective, i.e. X-ray photons are confined inside a sinking region due to large optical thickness of the flow \citep{2018MNRAS.476.2867M}. 
Under this condition, the temperature of plasma can be as high as a few hundred keV, which is sufficient to cause intense creation of electron-positron pairs \citep*{2019MNRAS.485L.131M} and further neutrino emission due to their annihilation \citep{1992PhRvD..46.3256K}. 
Thus, bright XRPs can manifest themselves as sources of intense neutrino emission,
where the total {energy flux released due to accretion} is channelled into luminosity in photons and luminosity in neutrinos.
The intrinsic neutrino luminosity of a NS is maximal right after the supernova explosion and then decreases rapidly with time: it is expected to be $<10^{37}\,\ergs$ ($<10^{36}\,\ergs$) after a $\mathrm{few}\times 10^2$\,years ($10^3$\,years) after the explosion \citep{2005NuPhA.752..590Y}, which is well below the expected neutrino luminosity in bright ULXs.

The higher the mass accretion rate and the total luminosity, the larger the fraction of energy released in the form of neutrinos (see Fig.\,1 in \citealt{2023MNRAS.522.3405A}). 
%AP5: \red{The highest neutrino luminosity is expected in ULX pulsars and bright Be X-ray transients \citep{2011Ap&SS.332....1R} if their apparent photon luminosity closely reflects the actual value (and therefore, the level of mass accretion rate), i.e., if the apparent luminosity is not significantly affected by X-ray collimation by the outflows launched from accretion disc. 
% For contrasting perspectives, see, e.g., \citealt*{2017MNRAS.468L..59K}, where the authors argue that this is not necessarily the case and photon luminosity in the brightest XRPs is drammatically affected by the outflows, and \citealt{2021MNRAS.501.2424M,2023MNRAS.518.5457M}, which present an opposing viewpoint.}
The highest accretion rates (and consequently neutrino luminosities) are expected in ULX pulsars and bright Be X-ray transients \citep{2011Ap&SS.332....1R}
because of their high apparent photon luminosities.
However, a high apparent luminosity may not necessarily correspond to a high accretion rate, if radiation is strongly collimated 
by accretion disc winds, as suggested for these objects by \citet*{2017MNRAS.468L..59K}; see also \citet{LasotaKing23}.
On the other hand, this hypothesis is still under debate; the arguments that disfavor strong 
collimation have been presented, e.g., by \citet{2021MNRAS.501.2424M,2023MNRAS.518.5457M}.

Due to the extra-galactic nature of confirmed ULX pulsars, the expected neutrino flux is very low and even expected to be significantly below the neutrino isotropic background in a MeV energy band \citep{2023MNRAS.522.3405A}. 
However,
%AP5: recent 
estimations of neutrino flux from ULX pulsars
\citep{2023MNRAS.522.3405A}
were performed under the assumption of isotropic neutrino emission, 
which can be violated by initial non-isotropic emission as well as by gravitational bending of particle trajectories.

In this paper, we investigate
neutrino beam pattern formation accounting for
neutrino propagation in a space curved by the gravity of a
star and transparency of a star
for neutrino emission.
The gravitational bending results
in a difference between actual
(initially generated)
$L_\nu$
and apparent $L_{\nu,\rm app}$ neutrino
luminosities.
The latter can be  different for different distant observers.
On the base of calculated beam patterns we
obtain
expected distributions of neutrino pulsars over the amplification factor
\beq \label{eq:a_nu}
a_\nu \equiv \frac{L_{\nu,\rm app}}{L_{\nu}}.
\eeq

%%%%%%%%%%%%%%%%%%%%%%%%%%%%%%%%%%%%%%%%
\section{Model setup}
\label{sec:Model}
%%%%%%%%%%%%%%%%%%%%%%%%%%%%%%%%%%%%%%%%

%%%%%%%%%%%%%%%%%%%%%%%%%%%%%%%%%%%%%%%%%%%
\subsection{Equation of state and structure of a neutron star}
\label{sec:NS_EoS}
%%%%%%%%%%%%%%%%%%%%%%%%%%%%%%%%%%%%%%%%%%%

We assume the NS structure to be spherical. Appreciable
deviations from the spherical symmetry can be caused by ultra-strong
magnetic fields ($B\gtrsim10^{17}$~G) or by rotation with ultra-short
periods (less than a few milliseconds;
{see, e.g., \citealt*{2007ASSL..326.....H} and references therein}),
but we will not consider such
cases.

The general static isotropic metric
satisfying the Einstein field equations can be written in the ``standard form''
using the Schwarzschild coordinates as follows
(see, e.g., section 8.1 in \citet{Weinberg:1972kfs} and section 23 in \citealt*{1973grav.book.....M}):
\beq
\label{eq:Sph_sym_metr}
\mathrm{d} s^2 = - B(r) \mathrm{d} t^2 
+ A(r)\mathrm{d} r^2
+ r^2(\mathrm{d}\theta^2+\sin^2\theta\,\mathrm{d}\phi^2),
\eeq
where $t$ is the time coordinate, $r$, $\theta$ and $\phi$ are the
spherical polar coordinates, 
\beq
   A(r) = \left( 1 - \frac{2GM_r}{c^2 r} \right)^{-1},
\quad 
   B(r) = e^{2\Phi/c^2},
\label{eq:AB}
\eeq
$M_r$ is the gravitational mass inside a sphere
of radius $r$,
$G$ is the Newtonian constant of gravitation, $\Phi$ is the
gravitational potential and $c$ is the speed of light in vacuum. Then
the mechanical structure of a NS is governed by four first-order
differential equations for $r$, $M_r$, $\Phi$ and the local pressure $P$
as functions of baryon number $\beta$ inside a given spherical shell 
(e.g., \citealp*{1979ApJS...39...29R}):
\begin{eqnarray}
  \frac{\mathrm{d} r}{\mathrm{d} \beta} &=& \frac{1}{4\pi r^2 \bar{n}\,\sqrt{A(r)}},
\label{drda}
\\
   \frac{\mathrm{d} M_r}{\mathrm{d} \beta} &=& \frac{\rho}{\bar{n}\,\sqrt{A(r)}},
\\
   \frac{\mathrm{d}\Phi}{\mathrm{d} \beta} &=& G\,\frac{M_r+4\pi r^3 P/c^2}{
         4\pi r^4\bar{n}}\,\sqrt{A(r)},
\label{Phi}
\\
   \frac{\mathrm{d} P}{\mathrm{d} \beta}&=& -\left(\rho+\frac{P}{c^2}\right)
         \,\frac{\mathrm{d}\Phi}{\mathrm{d} \beta},
\label{dPda}
\end{eqnarray}
where $\bar{n}$ is the mean number density of baryons.
We integrate these equations
from $r=0$ and $M_r=0$ at the
center of the star outwards, starting from a predefined baryon
density $\bar{n}$ at the center,
until a predefined mass density at the outer boundary
$\rho_\mathrm{b}$ is reached.
%AP5 Since the outermost non-degenerate layers are unimportant for the problem of neutrino propagation to be studied, we restrict ourselves by the strongly degenerate matter. This allows us to use a barotropic  equation of state (EoS), which provides mass density $\rho$ and pressure $P$ as functions of $\bar{n}$, regardless of temperature.

The boundary condition for the gravitational potential $\Phi$ is provided by the Schwarzschild metric outside the star,
\beq \label{PhiR}
   \mathrm{e}^{2{\Phi_\mathrm{b}/c^2}} = 1-\frac{2GM}{R c^2},
\eeq
where $R$ and $M=M_R$ are the stellar radius and mass, and
$\Phi_\mathrm{b}$ is the value of $\Phi$ at
the stellar surface. Since the value of $\Phi$ at the center of the NS
is not known in advance, we integrate equation~(\ref{Phi}) for a shifted
potential $\tilde{\Phi}(\beta)=\Phi(\beta)-\Phi(0)$, with the initial value
$\tilde{\Phi}(0)=0$ at the center of the star, and the value of the
shift $\Phi(0)$ is found from equation~(\ref{PhiR}) after the integration has
been completed.

We solve the set of equations (\ref{drda})\,--\,(\ref{dPda}) numerically
by the classic Runge-Kutta method on a non-uniform grid in $\beta$, with variable
steps adapted to provide a sufficient accuracy at each grid node 
for each of the computed functions. We decrease each step until a desired accuracy is reached.
In order to prevent accuracy loss in the outer
layers of the star,
where $\beta$ is nearly constant as a function of $\rho$ or
$r$, we use the difference $(\beta_\mathrm{b}-\beta)$ as an independent
variable,
$\beta_\mathrm{b}$ being the
%AP5: boundary value of $\beta$,
value of $\beta$ at $r=R$, which is
equal to the total number of baryons in the NS.

We limit ourselves to three EoSs: APR \citep*{1998PhRvC..58.1804A}, SLy4 \citep{2001A&A...380..151D} and BSk24 \citep{2018MNRAS.481.2994P}.
These EoSs describe the ground state for the nucleon-lepton ($npe\mu$) composition of matter, which is the most conservative assumption, without any ``exotic'' constituents.
The APR EoS for the NS core is based on realistic effective two- and three-nucleon interactions, which allow one to reproduce various nucleon scattering data and the properties of light nuclei.
We adopt the version of the APR EoS named A18+$\delta{v}$+UIX$^*$ in \citet{1998PhRvC..58.1804A}, which includes a relativistic boost correction, in the parametrized form of \citet{2018A&A...609A..74P}.
This parametrization includes the NS crust described by the BSk24 EoS on top of the core described by the APR EoS.
The SLy and BSk models are based on effective nucleon-nucleon interactions of the Skyrme type, adjusted to reproduce the EoS of pure neutron matter and the experimental properties of heavy atomic nuclei.
These EoS models are \emph{unified}: they are based on the same microscopic models for the core and the crust.
For the first (SLy) EoS family,
we use the SLy4 EoS in the parametrized form of \citet{2004A&A...428..191H}. 
The more recent BSk interaction model is more complicated; it is better tuned to the recent collection of experimental nuclear data. The BSk24 and BSk25 versions of this model, which are very similar, appear to be preferred, as discussed by \citet{2018MNRAS.481.2994P}. The BSk24 EoS is substantially stiffer than the SLy4 EoS, and we choose it as representative example of relatively stiff and soft EoS models.

%-----------------------------------------%
\begin{figure}
\centering 
\includegraphics[width=\columnwidth]{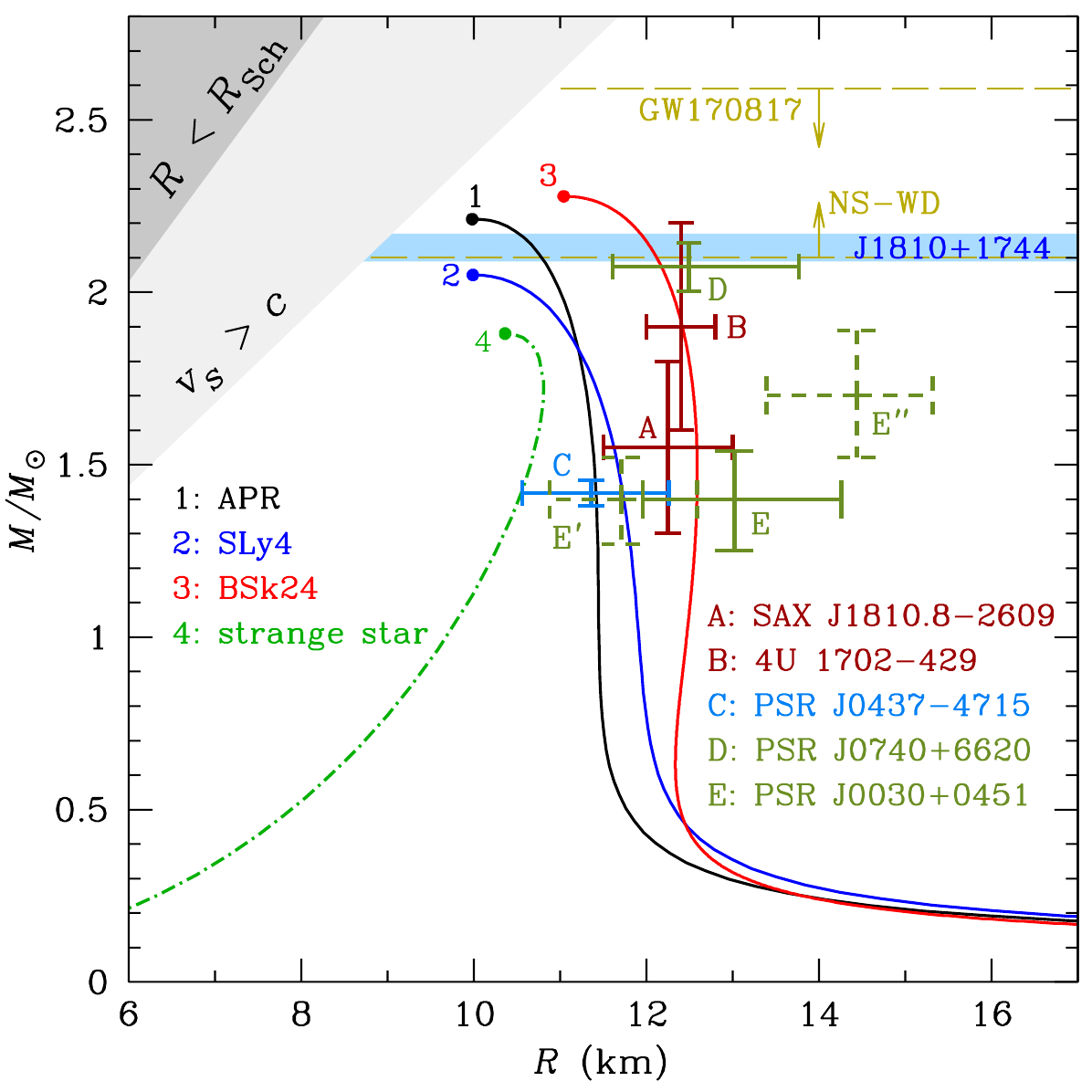}
\caption{NS mass $M$ versus radius $R$ for four EoS models compared with
theoretical and observational constraints and estimates.  The solid
curves labelled with numbers 1--3 show mass-radius relations for APR
(1), SLy4 (2) and BSk24 (3) EoS models, respectively. The dashed-dotted
line shows an example of mass-radius relation for a strange star. The
curves are truncated at the hydrostatic stability limits, marked by the
heavy dots. The dark and light grey shaded triangles are prohibited by
General Relativity and causality. The horizontal light-blue 
band corresponds to the mass estimate (with $1\sigma$ uncertainties) for
PSR~J1810+1744.
The horizontal dashed lines with arrows mark the upper and lower 3$\sigma$ limits to the maximum NS mass, derived from observations. 
The error bars show the 1$\sigma$ confidence intervals in $M$ and $R$, inferred from observations of five
NSs in binary systems, marked by the letters from A through E, according to the legend 
(see text for details).
}
\label{pic:NS_MR}
\end{figure}
%-----------------------------------------%

We also consider a possibility of the EoS of strange matter built
of the $u,\,d$ and $s$ quarks \citep*{1984PhRvD..30..272W,1986A&A...160..121H,1986ApJ...310..261A}. For the
strange matter EoS, we use the
approximation proposed by \citet{2000A&A...359..311Z}
and adopt the fiducial parameters in his paper: the bag constant $B=60$ MeV fm$^{-3}$, the QCD
coupling constant $\alpha_c=0$ and the rest energy of the strange quark $m_s c^2=100$~MeV.

The solid curves in Fig.~\ref{pic:NS_MR} show gravitational mass $M$ versus circumferential radius $R$ of a NS for the three selected EoSs.
The dot-dashed curve displays $M(R)$ for a strange star model.
The dark grey shaded
area ($R < R_\mathrm{Sch}$, where $R_\mathrm{Sch} = 2GM/c^2$ is the Schwarzschild radius) is prohibited by General
Relativity. The entire grey shaded triangle is prohibited by General
Relativity combined with the condition that the speed of sound must be
subluminal (e.g., section~6.5.7 of \citealt{2007ASSL..326.....H}).

The lower horizontal dashed line with the upward arrow marks the lower
3$\sigma$ limit to the maximum NS mass, $M_\mathrm{max} > 2.09$,
obtained by \citet{2022ApJ...934L..17R} jointly for seven most massive
known NSs in binaries with white dwarfs (WDs). The light-blue horizontal
band corresponds to the most accurate of individual estimates for these
NSs ($M=2.13\pm0.04\,M_\odot$ for PSR J1810+1744). These mass
estimates rely on an analysis of orbital light curves with a specific
model for heating of the WD surface by radiation from a pulsar, hence
they can be model-dependent. Pulsar mass estimates obtained using the
effects of General Relativity (in particular the Shapiro delay of the
pulsar signal) appear to be less model-dependent. However, one of the largest
estimates of this kind ($M=2.01\pm0.04\,M_\odot$ for PSR J0348+0432;
\citealt{2013Sci...340..448A}) was recently revised to a 10\% lower
value ($M=1.806\pm0.037\,M_\odot$; \citealt{Saffer_J0348}). The
highest reliable pulsar mass estimates, based on the Shapiro delay
measurements, are currently $M=2.073\pm0.069\,M_\odot$ with the
lower bound $M>1.95\,M_\odot$ at the  95.4\% confidence level for PSR J0740+6620 \citep{2021ApJ...915L..12F} and $M=1.908\pm0.016\,M_\odot$ for PSR
J1614$-$2230 \citep{2018ApJS..235...37A}.

\citet*{2018ApJ...852L..25R} used quasi-universal relations exhibited by equilibrium solutions of rotating relativistic stars to infer constraints on the maximum NS mass from an analysis of the electromagnetic and gravitational wave signals from the double NS merger GW170817. 
Their most conservative upper limit $M_\mathrm{max}<2.59\,M_\odot$ is shown in Fig.~\ref{pic:NS_MR} by the upper horizontal dashed line with the downward arrow. 
It relies on the assumption that the merger product in GW170817 has collapsed into a black hole.
However, the kilonova produced in this event could also be explained in an alternative scenario of NS stripping without black hole formation \citep{2022Parti...5..198B}. 
Thus the indicated limit is model-dependent.

%-----------------------------------------%
\begin{figure*}
\centering 
\includegraphics[width=16.cm]{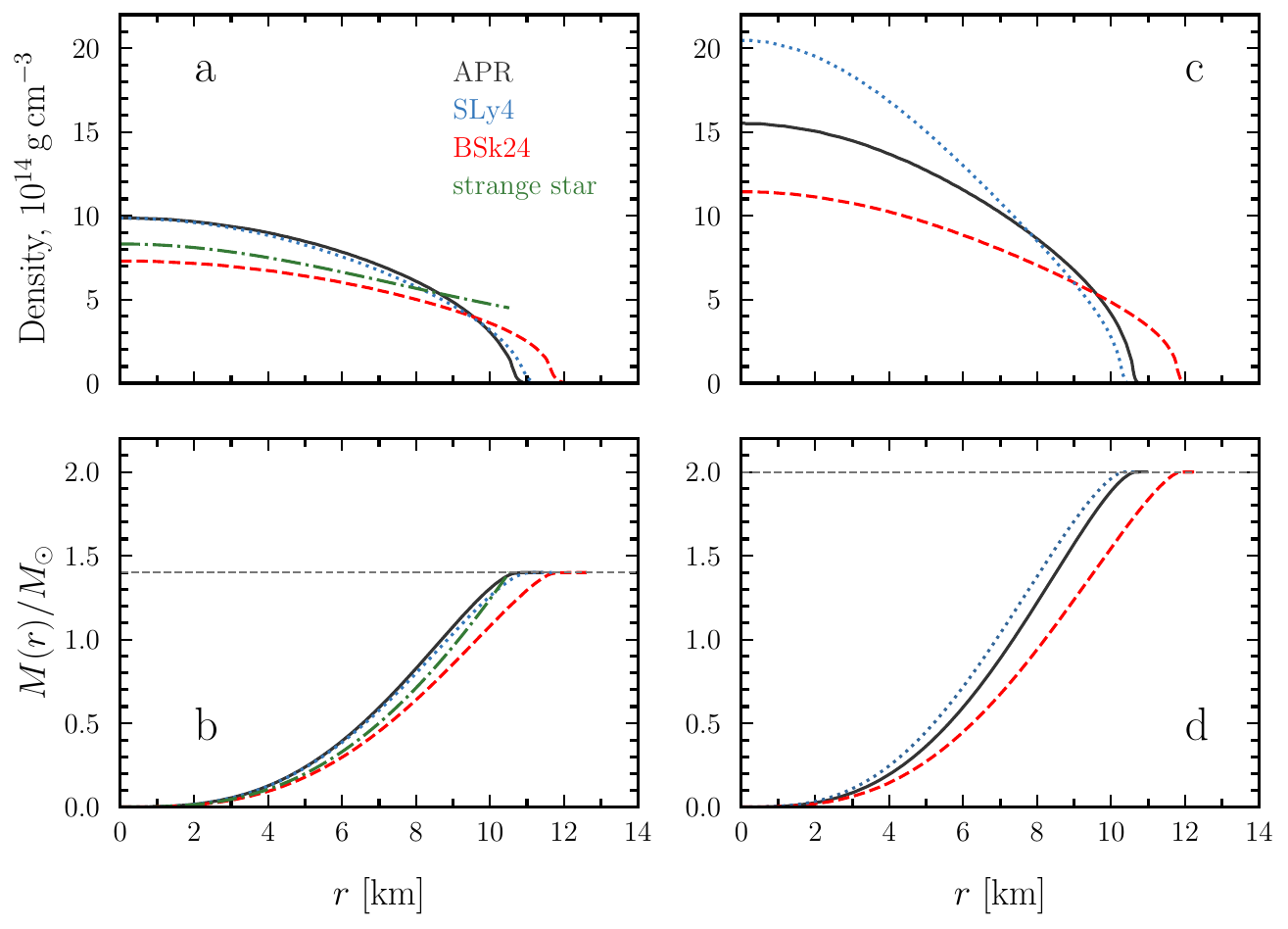} 
\caption{
Mass density (a,c) and mass (b,d) distributions inside a {NS/strange star} of a given total mass and EoS.
Different curves are calculated for different EoS:
APR (black solid), 
SLy4 (blue dotted),
BSk24 (red dashed), 
and EoS for a strange star according to \citet{2000A&A...359..311Z}.
Mass of a NS is fixed at $M=1.4 M_\odot$ at panels ``a'' and ``b'',
and at $M=2 M_\odot$ at panels ``c'' and ``d''.
Mass of a strange star was taken to be $M=1.4M_\odot$ at panel ``a''.
}
\label{pic:NS_structure}
\end{figure*}
%-----------------------------------------%

The vertical and horizontal error bars in Fig.~\ref{pic:NS_MR} show the available 1$\sigma$ confidence intervals in $M$ and $R$, respectively,
for the cases where these uncertainties are not large ($\la15$\%). 
The labels A and B mark such intervals for the bursters SAX~J1810.8--2609 and 4U~1702-429, according to the analysis by
\citet{2017A&A...608A..31N}. The label C corresponds to the nearest and brightest millisecond pulsar PSR~J0437--4715 in an NS-WD binary system, according to \citet{2024ApJ...971L..20C}. Labels D and E mark the
results obtained for PSR J0740+6620 \citep{2024ApJ...974..294S} and PSR
J0030+0451 \citep{2019ApJ...887L..24M}, respectively, using an analysis
of the energy-dependent thermal X-ray waveform observed by \emph{NICER}.
Despite the belief that this approach was ``less subject to systematic
errors than other approaches for estimating neutron star radii''  \citep{2019ApJ...887L..24M}, a subsequent reanalysis, performed for PSR J0030+0451 by
\citet{2024ApJ...961...62V} with alternative hot spot models and using jointly the data of \emph{NICER} and \emph{XMM-Newton}, resulted in substantially different estimates, shown in Fig.~\ref{pic:NS_MR} by the
dashed error bars and marked as E$'$ and E$''$, which demonstrate the strong model dependence. 
Other joint mass and radius estimates obtained from spectral analyses of X-ray radiation of
neutron stars exhibit similar model dependence and are not plotted here 
(e.g., \citealt{2022MNRAS.516...13T}; see also discussion and references
in \citealt{2020MNRAS.496.5052P}).

Fig.~\ref{pic:NS_MR} demonstrates that the selected EoS models are reasonably compatible with the available observational NS mass and radius estimates, although the softest SLy4 EoS is only marginally compatible with the lower limits to $M_\mathrm{max}$. 

Solutions of the stellar structure equations for the three selected NS
EoSs are illustrated in Fig.~\ref{pic:NS_structure}. 
The upper and lower panels show respectively mass density $\rho$ and gravitational mass distributions $M_r$ as functions of $r$ for the total mass of a NS equal to $1.4\,M_\odot$ (left panels) or $2\,M_\odot$ (right panels).
The dot-dashed line in the left panels show analogous distributions for the model of a strange star with $M=1.4\,M_\odot$.

%%%%%%%%%%%%%%%%%%%%%%%%%%%%%%%%%%%%%%%%%%%%%%
\subsection{Neutrino emission and propagation}
\label{sec:neutrino_emission_propagation}
%%%%%%%%%%%%%%%%%%%%%%%%%%%%%%%%%%%%%%%%%%%%%%
\subsubsection{Neutrino opacities}
\label{sec:opac}

The distinct feature of neutrino propagation near a NS stems from the fact that NSs are typically transparent to neutrino emissions of relatively low energy \citep{1979ApJ...230..859S,1987A&A...179..127H},
unless their internal temperature $T$ exceeds
%AP5: $\sim 1\,{\rm MeV}$ ($\sim10^{10}$~K),
$\sim10^{10}$~K ($k_\mathrm{B}T\sim 1$~MeV, where $k_\mathrm{B}$ is the Boltzmann constant), which occurs only immediately following a supernova explosion (see, e.g., Figs.~3 and 9 in \citealt{2018A&A...609A..74P}).
In bright XRPs, it is expected that neutrinos are emitted near the base of the accretion column.
We assume that the magnetic field near the NS surface is dominated by the dipole component, leading to neutrinos being initially emitted near the magnetic poles of the NS located diametrically opposite 
on the stellar surface.\footnote{Note that in a few XRPs, non-dipole magnetic field structures have been proposed to explain observational data (see, e.g., \citealt{2013MNRAS.435.1147P,2017A&A...605A..39T,2017Sci...355..817I,2022MNRAS.515..571M}).}
The primary process of neutrino production involves the annihilation of electron-positron pairs, although a fraction of neutrinos can also be generated via the synchrotron process (see, e.g., \citealt{1992PhRvD..46.3256K} and Mushtukov et al., in prep.).
Emitted neutrinos propagate both outside and inside the NS, which is cold enough to be nearly transparent to them.
All emitted neutrinos undergo gravitational bending in curved space-time.
Neutrino emission and their subsequent propagation along curved trajectories form a distinct beam pattern.
While the vast majority of neutrinos initially emitted at the NS surface are of electron flavor, the composition of the neutrino flux varies as it propagates, because of neutrino oscillations.
Neutrinos of very high energies $E_\nu$ (above several MeV; see below) may experience scatterings in the core of the NS, which influence their trajectories and, consequently, their final angular distribution, but we will not consider such high energies here.

The primary processes
governing neutrino opacity in dense matter are neutrino-neutron scattering (see, e.g., \citealt{Shapiro:1983du})
%AP6:
%\beq\label{eq:neutrino-neutron_scattering}
%n + \nu_{e,\mu}\,\longrightarrow \,n + \nu_{e,\mu}.
%\eeq
%and neutrino absorption
%\beq\label{eq:neutrino_abs}
%n + \nu_{e}\,\longrightarrow p + e^-,
%\qquad
%n + \nu_{\mu}\,\longrightarrow p + \mu^-.
%\eeq
% AK
%These processes are strongly suppressed in the case of relatively low temperatures
%because of the degeneracy of the nucleons
%\green{
%The latter processes are strongly suppressed in the degenerate NS matter at $p_\mathrm{Fn} \gg p_\mathrm{Fp} + p_\mathrm{Fe}$ 
%(see, e.g., \citealt{1979ApJ...230..859S,1987A&A...179..127H})
%for the same reason as the direct Urca process  
\beq\label{eq:neutrino-neutron_scattering}
n + \nu_{e,\mu}\,\longrightarrow \,n + \nu_{e,\mu}.
\eeq
and neutrino absorption
\beq\label{eq:neutrino_abs}
n + \nu_{e}\,\longrightarrow p + e^-,
\qquad
n + \nu_{\mu}\,\longrightarrow p + \mu^-.
\eeq
%(see, e.g., section 11.2 in \citealt{Shapiro:1983du}), 
%where $p_\mathrm{F(n,p,e)}$ are the Fermi momenta of neutrons, protons and electrons, respectively.
The neutrino mean free path in the elastic limit of neutrino-neutron scattering (\ref{eq:neutrino-neutron_scattering})
for non-degenerate nucleons can be estimated as 
\citep[e.g., equation (11.7.8) in][]{Shapiro:1983du}
\beq\label{eq:scattering-non-degenerate}
\overline{\lambda}_\mathrm{sc,0} \sim
3 \, \frac{\rho_\mathrm{nuc}}{\rho} \left( \frac{\mbox{1~MeV}}{E_{\nu}} \right)^{\!2} \mbox{~km} ,
\eeq
where $\rho_\mathrm{nuc} \approx 2.5 \times 10^{14}\,{\rm g\,cm^{-3}}$ denotes the mass density at the saturation number density of baryons $n_\mathrm{nuc} \approx 0.15$ fm$^{-3}$ \citep*{2020PhRvC.102d4321H}.

However, at temperatures $T\lesssim 10^8$~K
and densities $\rho\sim(1$--$5)\rho_\mathrm{nuc}$, typical for mature NSs (unlike, for instance, non-degenerate regions of a proto-NS), the neutrino energies most relevant for the ULXs ($E_\nu\sim0.1$--1~MeV) are large compared with temperature $T$
but small compared with neutron Fermi energy
$E_{\mathrm{F n}}$.
% \green{Thus in the bulk of a typical NS
% {absorption of neutrinos proceeds 
% (in analogy with the modified Urca processes) 
% via the  reactions
% \beq\label{eq:neutrino-mod_abs}
% n + n + \nu_{e}\,\longrightarrow p + n + e^-,
% \quad
% n + p + \nu_{e}\,\longrightarrow p + p + e^-,
% \eeq
% \citep{1979ApJ...230..859S,Iwamoto82,1987A&A...179..127H}
%(see, e.g., \citealt{1979ApJ...230..859S,1987A&A...179..127H}).
%AP6: The 
Under these conditions, the mean free path $\overline{\lambda}_\mathrm{sc}$
increases compared with $\overline{\lambda}_\mathrm{sc,0}$ by a factor $\propto E_{\mathrm{F n}}/E_\nu$ 
(see, e.g., equation (7.2) in \citealt{Iwamoto82}). Then the neutrino mean free path 
$\overline{\lambda}_\mathrm{sc}$
in the elastic limit of neutrino-neutron scattering (\ref{eq:neutrino-neutron_scattering}) 
%{for non-degenerate nucleons}
%can be estimated as 
%\citep[e.g., equation (11.7.8) in][]{Shapiro:1983du}
%\beq\label{eq:lambda_sc}
%\overline{\lambda}_\mathrm{sc,0} \sim
%3 \, \frac{\rho_\mathrm{nuc}}{\rho} \left( \frac{\mbox{1~MeV}}{E_{\nu}} \right)^{\!2} \mbox{~km} ,
%\eeq
%where $\rho_\mathrm{nuc} \approx 2.5 \times 10^{14}\,{\rm g\,cm^{-3}}$ 
%denotes the mass density at the saturation number density of baryons $n_\mathrm{nuc} \approx 0.15$ %fm$^{-3}$ \citep*{2020PhRvC.102d4321H}.
%Since the mass density $\rho$ in the NS core substantially exceeds $\rho_\mathrm{nuc}$, the mean free path of high-energy neutrinos may be smaller than the radius of a NS.
%Consequently, an appreciable fraction of neutrinos may undergo scatterings on non-degenerate nucleons in a very hot NS (a proto-NS)
%However, 
%at temperatures
%AP5: $\lesssim 10\,{\rm keV}$
%$T\lesssim 10^8$~K and
%densities $\rho\sim(1-5)\rho_\mathrm{nuc}$, typical for mature NSs, 
%and the neutrino energies $E_\nu\sim0.1-1$~MeV
% \green{(most relevant for the ULXs)
%($E_\nu\sim0.1-1$~MeV) are 
%that are }
%large compared to temperature $T$ but small compared to  neutron Fermi energy 
%AP5:
%\purple{$E_{\mathrm{F} \nu}$. 
%\green{Under these conditions, the mean free path $\overline{\lambda}_\mathrm{sc}$
%increases compared with $\overline{\lambda}_\mathrm{sc,0}$ by a factor $\propto E_{\mathrm{F}\nu}/E_\nu$ (see, e.g., equation (7.2) in \citealt{Iwamoto82}).
%At $\rho\approx\rho_\mathrm{nuc}$ it}
can be estimated from equation (26) in \citet{1979ApJ...230..859S} or 
(similar  assessment)
equation (7.2) in \citet{Iwamoto82},
\beq\label{eq:lambda_sc} 
\overline{\lambda}_\mathrm{sc} \approx 800\ \left(\frac{n_\mathrm{nuc}}{n_\mathrm{n}}\right)^{\! 2/3} \left(\frac{\mbox{1~MeV}}{E_\nu}\right)^{\!3} \mbox{~km},
\eeq 
where $n_\mathrm{n}$ is the number density of neutrons,
so that $\overline{\lambda}_\mathrm{sc}\gg R$.

%AP6:
The neutrino absorption process (\ref{eq:neutrino_abs}) is forbidden in the degenerate NS matter at $p_\mathrm{Fn} \gg p_\mathrm{Fp} + p_\mathrm{Fe}$ (for the same reason as the direct Urca process; cf., e.g., section 11.2 in \citealt{Shapiro:1983du}), where $p_\mathrm{F(n,p,e)}$ are the Fermi momenta of neutrons, protons and electrons, respectively.
Thus in the bulk of a typical NS, absorption of neutrinos proceeds (in analogy with the modified Urca processes) via the modified reactions
\beq
n + n + \nu_{e}\,\longrightarrow p + n + e^-,
\quad
n + p + \nu_{e}\,\longrightarrow p + p + e^-,
\eeq
with the mean free paths  $\overline{\lambda}_\mathrm{abs}$ still longer than $\overline{\lambda}_\mathrm{sc}$ at the low temperatures 
%AP5: [REWRITTEN. THIS PART HAS BEEN DOUBLE-CHECKED AND CORRECTED]
% \citep[cf.][]{1987A&A...179..127H}. In this case
% we can estimate the neutrino mean free path before absorption, $\overline{\lambda}_\mathrm{abs}$,
% using equations~(23) and (25) in \citet{1987A&A...179..127H}. 
% For example, at $E_\nu\sim100$~keV, $\rho\sim10^{15}$ g~cm$^{-3}$ and temperature $\sim10^8$~K
% we obtain
% $\overline{\lambda}_\mathrm{abs} \sim 10^5$~km.
% Equation~(16) in \citet{1979ApJ...230..859S} gives a similar result.
\citep{1979ApJ...230..859S,Iwamoto82,1987A&A...179..127H}.
In this case $\overline{\lambda}_\mathrm{abs}$ is given by equation (16) in \citet{1979ApJ...230..859S} or equation (20) in \citet{1987A&A...179..127H}. 
At the nuclear saturation density, we can estimate $\overline{\lambda}_\mathrm{abs}$ using the convenient formula (22) in \citet{1979ApJ...230..859S}, which agrees with figure~1 in \citet{1987A&A...179..127H} at $T=5\times10^{10}$~K and reproduces the scaling law (25) in \citet{1987A&A...179..127H}. 
At the densities and temperatures typical for the core of a mature NS, $\overline{\lambda}_\mathrm{abs}$ increases with decreasing energy $E_\nu$ from $\overline{\lambda}_\mathrm{abs}\sim(1-3)\times10^6$ km at $E_\nu=1$ MeV to $\overline{\lambda}_\mathrm{abs}\sim(1-3)\times10^{10}$ km at $E_\nu=0.1$ MeV.

%AP5: If the NS core contains quark matter, 
In the quark (degenerate)  matter,
in contrast to the  nucleon  matter,
neutrino scattering is less efficient than absorption by quarks (e.g., \citealt{Pal:2011ve}),
because the processes similar  to  (\ref{eq:neutrino_abs}), e.g.,\
$d + \nu_e \to u + e^-$,   are allowed  (see, e.g., \citealt{Iwamoto82}) .   
In this case we have
$\overline{\lambda}_\mathrm{qabs} / \overline{\lambda}_\mathrm{qsc} \ll 1$, 
where $\overline{\lambda}_\mathrm{qabs}$ and $\overline{\lambda}_\mathrm{qsc}$ 
are electron neutrino mean free paths due to absorption and scattering in quark matter respectively.
%AP5:
%The electron neutrino mean free path due to the absorption {for $E_\nu\sim 500\,{\rm keV}$} can be estimated as
%\beq\label{eq:free_path_QS}
%\overline{\lambda}_\mathrm{qabs} \approx
%\frac{\pi^4}{4 \, \alpha_s \cos^2\theta_c} \frac{1}{G_\mathrm{F}^2 \mu_d \mu_u \mu_e E_\nu^2}
%\approx {56 \, {\rm km},}
%\eeq
%where 
%$G_\mathrm{F}$ is the Fermi weak-coupling constant,
%$\theta_c$ is the Cabibbo angle,
%and the numerical estimation is based on the typical parameters for quarks in the core of a compact star $\mu_d \approx \mu_u \approx 500$~MeV, %$\mu_e \approx 11$~MeV 
%and $\alpha_s \approx 1$ \citep{Schafer:2004jp}.
%Based on equation~(\ref{eq:free_path_QS}), we use the estimate
%\beq 
%\overline{\lambda}_\mathrm{qabs} \sim 
%14\,\left(\frac{\mbox{1~MeV}}{E_\nu}\right)^{\!2}\mbox{~km}.
%\eeq

The absorption coefficient of non-degenerate neutrinos in the quark matter can be assessed with equation~(30) by \citet{Pal:2011ve}
%AP6: \green{ (slightly corrects  equation~(6.8) of \citealt{Iwamoto82}).}  
Let us assume $E_\nu\gg \pi k_\mathrm{B}T$,
adopt the standard quark color factor $C_F=4/3$ and neglect higher-order corrections in this equation, retaining only the leading term, which is equivalent to equation~(6.8) of \citet{Iwamoto82}.
% and adopt the standard quark color factor $C_F=4/3$ in that equation.
Then we have
\beq\label{eq:free_path_QS}
\overline{\lambda}_\mathrm{qabs} \approx \frac{\text{14 km}}{\alpha_\mathrm{s}\mu_{u,500}\mu_{d,500}\mu_\mathrm{e,11}}
\left(\frac{\text{1 MeV}}{E_\nu}\right)^2,
\eeq
where $\alpha_\mathrm{s}$ is the strong coupling constant, $\mu_{u,500}=\mu_u/(500$ MeV), $\mu_{d,500}=\mu_d/(500$ MeV), $\mu_\mathrm{e,11}=\mu_\mathrm{e}/(11$ MeV), $\mu_{u,d,\mathrm{e}}$ being the chemical potentials of the $u$ and $d$ quarks and of the electrons.
For estimates, following \citet{Schafer:2004jp} and \citet{Pal:2011ve}, we set $\alpha_\mathrm{s}=\mu_{u,500}=\mu_{d,500}=\mu_\mathrm{e,11}=1$, which corresponds to densities $\rho\approx6\rho_\mathrm{nuc}$.

%%%%%%%%%%%%%%%%%%%%%%%%%%%%%%%%%%%%%%%%%%%%%%
\subsubsection{Neutrino trajectories}
\label{sec:traj}

To characterize the geometry of spacetime around a NS, we employ the 
static spherically symmetric metric in the standard form, equation~(\ref{eq:Sph_sym_metr}).
It is a suitable approximation for NSs in XRPs with typical spin periods $P_\mathrm{spin} \ga 0.1\,{\rm s}$ \citep[see, e.g.,][]{2007ASSL..326.....H}.
For a spherically symmetric NS in a hydrostatic equilibrium, metric (\ref{eq:Sph_sym_metr}) is locally similar to the Schwarzschild metric produced by mass $M_r$.
In this metric,
%AP5: trajectories of photons and neutrinos between potential scattering events lie in 
each trajectory of a freely propagating neutrino lies in one and
the same plane.
Within
%AP5: the particle trajectory plane, 
this plane,
we parameterize the trajectory using polar coordinates, with $r\geq 0$ and $\varphi\in[0,2\pi]$. Since the neutrino rest masses are negligibly small compared with the considered neutrino energies, we describe a neutrino trajectory by an equation for a particle with zero rest mass.

The derivation of the general equations of motion in metric (\ref{eq:Sph_sym_metr}) can be found, e.g., in \citet{Weinberg:1972kfs}.
In particular, a trajectory of a particle in the equatorial plane (i.e.,
with the polar coordinate $\theta=\pi/2$) is given by equation (8.4.29) of \citet{Weinberg:1972kfs}.
For a massless particle, we should set the right-hand side of this equation to zero, which 
leads to 
\beq
    \frac{A(r)}{r^4}\left( \frac{\d r}{\d\varphi} \right)^{\!2} 
    + \frac{1}{r^2}
    = \frac{1}{B(r)b^2},
\label{eq:DE_traj}
\eeq
where $A(r)$ and $B(r)$ are determined by equation~(\ref{eq:AB}),
and  $b$   
%is a constant corresponding to the Killing vector $\partial/\partial\varphi$.  
%The  parameter $b$ 
has the physical meaning of the impact parameter,
which is constant along every given trajectory due to the angular momentum conservation. 
For each trajectory, $b$ is determined by the initial neutrino direction.
At the NS surface, it is related to the angle $\zeta$ between the radial direction and neutrino trajectory as (see Appendix~\ref{app:GR_DU})
\beq\label{eq:beta2zeta} 
 b = \frac{R \sin\zeta}{\sqrt{1 -  R_\mathrm{Sch}/R}},
\eeq
where $R_\mathrm{Sch} = 2 G M /c^2$ is the Schwarzshild  radius.

In the empty space outside a NS (at $r>R$), $A(r)B(r)=1$ (see, e.g., \citealt{Weinberg:1972kfs}, section 8.2).
In this case, equation~(\ref{eq:DE_traj}) reduces to equation~(25.55) in \citet{1973grav.book.....M}, which
describes a photon trajectory in the Schwarzschild metric.
Inside a NS (at $r<R$), the functions $A(r)$ and $B(r)$ are determined by the EoS through 
the solution of the hydrostatic equilibrium equations (\ref{drda})--(\ref{dPda}).
Consequently, trajectories of neutrinos propagating through a NS are contingent upon the mass distribution within the star and are thus anticipated to vary for different EoSs.

%%%%%%%%%%%%%%%%%%%%%%%%%%%%%%%%%%%%%%%%%%%
\subsection{Neutron star rotation and luminosity distribution}
%%%%%%%%%%%%%%%%%%%%%%%%%%%%%%%%%%%%%%%%%%%

%AP5: [I HAVE REPLACED TIME INTERVAL NOTATION T TO t_\mathrm{av}]
The apparent luminosity of a NS can be determined as 
\beq \label{eq:L_ave}
L_\mathrm{\nu, app}=\frac{4\pi D^2}{t_\mathrm{av}}
\int\limits_{0}^{t_\mathrm{av}} F_\nu(t)\d t,
\eeq 
where $F_\nu(t)$ 
is variable neutrino energy flux density (as registered by a distant observer), 
which varies with time $t$, $D$ is a distance to the compact object and $t_\mathrm{av}$ is a time interval
%AP5:
for the averaging.
In practice, the integration in (\ref{eq:L_ave}) is performed over a long time interval ($t_\mathrm{av}\gg P_\mathrm{spin}$)
because the mass accretion rate in X-ray binaries is known to be fluctuating over a wide range of time scales, which should result in fluctuating pulse profiles in X-rays and neutrinos.
The ratio of the apparent and actual neutrino luminosity determines the neutrino amplification factor $a_\nu$,  equation~(\ref{eq:a_nu}).

Apparent neutrino luminosity depends on actual neutrino luminosity, neutrino beam pattern and geometry of NS rotation in the observer's reference frame.
Rotation of a NS in the observer's reference frame is described by two angles: inclination $i$ (i.e., the angle between the rotation axis and observer's line of sight) and the magnetic obliquity $\theta_B$ (i.e., the angle between the rotational and magnetic axis of a NS).
The flux is related to the neutrino flux distribution in the reference frame of a NS,
which depends on the angle $\psi$
between the observer's line of sight and NS magnetic axis at a given phase $\varphi_\mathrm{p}\in[0;2\pi]$ of NS rotation:
\beq \label{eq:psi}
\cos\psi = \cos i\cos\theta_B + \sin i\sin \theta_B\cos\varphi_\mathrm{p}.
\eeq
The angles $i$ and $\theta_B$ are typically unknown for the XRPs.
Recent observation of X-ray polarization variable over NS spin period, however, shed light on rotation geometry in some particular accreting strongly magnetized NSs \citep{2022NatAs...6.1433D,2022ApJ...941L..14T,2023A&A...675A..48T,2023A&A...677A..57D,2023MNRAS.524.2004M,2023A&A...675A..29M,2023arXiv231103667H,2024A&A...691A.216F}, but features of NS distribution over rotation parameters are still uncertain. 
To estimate possible deviations of apparent neutrino luminosity from the actual one, we assume a random distribution of NSs over
%AP5: the parameters of their rotation, 
the angles $i$ and $\theta_B$,
simulate neutrino pulse profiles for various
%AP5: rotation parameters
$i$ and $\theta_B$,
and calculate theoretical distributions $f(a_\nu)$ of NSs over the apparent neutrino luminosity amplification factors $a_\nu$.
The technique used here is similar to the one applied to investigate distributions of XRPs over the apparent luminosity in X-rays (see, e.g. \citealt{2021MNRAS.501.2424M,2024MNRAS.527.5374M}).

%%%%%%%%%%%%%%%%%%%%%%%%%%%%%%%%%%%%%%%%
\section{Numerical model}
\label{sec:NumModel}
%%%%%%%%%%%%%%%%%%%%%%%%%%%%%%%%%%%%%%%%

Our numerical
%AP5: model consists of two parts.
procedure consists of two stages.
First, we compute the angular distribution of neutrino flux in the reference frame of a NS accounting for neutrino propagation along curved trajectories and
%AP5: scattering/
absorption inside a star.
Then, using the computed angular distribution, we simulate neutrino flux variability in the observer's reference frame due to the rotation of a NS and calculate the theoretical distribution of neutrino pulsars over the neutrino amplification factor. 

%%%%%%%%%%%%%%%%%%%%%%%%%%%%%%%%%%%%%%%%%%%
\subsection{Neutrino angular distribution}
\label{sec:Num_ang_dist}
%%%%%%%%%%%%%%%%%%%%%%%%%%%%%%%%%%%%%%%%%%%

To obtain the angular distribution of neutrinos, we specify neutrino energy $E_\nu$, 
NS EoS and mass, which gives us NS radius and internal mass distribution. 
Then we perform Monte Carlo simulations
%AP5: and
to
calculate trajectories of $4\times 10^7$ particles in each run.

%AP5: There are a few steps in our Monte Carlo simulation:
As an example, we consider the case when neutrino absorption is much more efficient than scattering,
so scattering can be neglected.
Then the simulation includes the following steps.
\begin{enumerate}[leftmargin=12pt,wide]
\item\label{step:neutrino_ini}
We start with neutrino of energy $E_\nu$ emitted from the surface of a NS near one of its magnetic poles.
The initial direction of particle motion is taken to be random and calculated under 
the assumption that the initial angular distribution of neutrinos is isotropic. 
\item \label{step:free_path}
We choose a random realization of the optical depth traveled by the particle before
%AP5: the absorption event
absorption  (the dimensionless free path) according to the formula
$
\tau_X = -\ln X,
$
where $X\in(0;1)$ is a random number having the uniform distribution.
%Neutrino scattering 
%is  neglected
%%AP5: as we limit ourselves by consideration of relatively cold NS and strange stars.
%according to Section~\ref{sec:opac}.
\item
We simulate a particle trajectory by solving numerically differential equation (\ref{eq:DE_traj}) for a set of 
initial parameters with initial impact factors calculated according to equation~(\ref{eq:beta2zeta}) (see Appendix\,\ref{app:TrNum}).
If the trajectory crosses the star, we calculate an optical depth traveled by the particle, $\tau_\nu$, 
by integration along the simulated trajectory,
\beq\label{eq:tau} 
\tau_\nu(s) = \int\limits_0^s \frac{\d s'}{
{\overline{\lambda}}(s')},
\eeq 
where $s$ is the length along the particle trajectory and
\beq
\overline{\lambda}=\frac{1}{
\overline{\lambda}_\mathrm{abs}^{-1}+\overline{\lambda}_\mathrm{sc}^{-1}}
\simeq \overline{\lambda}_\mathrm{abs}
\eeq
is  the mean free path accounting for absorption 
($\overline{\lambda}_\mathrm{abs}$)  and  scattering ($\overline{\lambda}_\mathrm{sc}$), 
which depend on neutrino energy $E_\nu$ and mass density $\rho$ at each given point along the path inside 
the star according to the estimates in Section\,\ref{sec:neutrino_emission_propagation}.
%AP5:
The length element $\d s'$ in equation~(\ref{eq:tau}) is calculated according to equation~(\ref{eq:ds}) in Appendix~\ref{app:GR_DU}.

The optical depth (\ref{eq:tau}) is a non-decreasing function, 
bounded from above by some maximum value for every simulated trajectory. 
If $\tau_X$ exceeds this maximum, the particle goes to infinity without 
absorption 
and we account for its final momentum direction in the simulated angular distribution function. 
Then we return to step \ref{step:neutrino_ini} and start the simulation for the next particle. 
Otherwise, $\tau_\nu(s)$ reaches $\tau_X$ at some point of the trajectory.
In this case, the particle is considered to be absorbed,
we drop it from further consideration, return to step \ref{step:neutrino_ini} 
and start the simulation for the next neutrino.  Thus
simulating trajectories of a large number of particles, we arrive at the final angular 
distribution of neutrinos in the NS reference frame.
\end {enumerate}
%AP5: Note that the developed algorithm would be most relevant for hot NSs, that are not transparent to neutrino emission due to the scattering and absorption. In our case, however, the scattering and absorption processes do not noticeably affect the final angular distributions of neutrino for the NSs (see Section\,\ref{sec:neutrino_emission_propagation}) and are considered rather for the sake of generality.

The described algorithm is rather general.
However, as argued in Section\,\ref{sec:opac},   the scattering  and  absorption 
do not noticeably affect neutrino flux 
and can be neglected in a cold NS at the energies considered in the present study. 
On the other side,
the neutrino absorption can play a noticeable role in a strange quark star, 
as will be seen below in Figures~\ref{pic:sc_NS_nu_flux_real_QS}.

%%%%%%%%%%%%%%%%%%%%%%%%%%%%%%%%%%%%%%%%%%%
\subsection{Neutrino amplification factor}
\label{sec:Num_amp_fac}
%%%%%%%%%%%%%%%%%%%%%%%%%%%%%%%%%%%%%%%%%%%

To get the amplification factor (\ref{eq:a_nu}) for given rotation parameters $i$ and $\theta_B$, we use pre-calculated neutrino flux distribution in the reference frame of a star (Section\,\ref{sec:Num_ang_dist}) and apply equation~(\ref{eq:psi}) to compute the theoretical pulse profile in neutrino emission.
In our simulations, the mass accretion rate is assumed to be constant.
Under this condition, the pulse profile does not experience variations from one pulse period to another, and we use $t_\mathrm{av}=P_\mathrm{spin}$ in equation (\ref{eq:L_ave}).
Averaging the neutrino flux variable over the NS spin period, we obtain the apparent neutrino luminosity (\ref{eq:L_ave}). 
Dividing it by the actual neutrino luminosity $L_\nu$, defined as the initial neutrino flux integrated over the emission region as seen by a distant observer (that is, corrected for the gravitational redshift), we obtain the neutrino amplification factor (\ref{eq:a_nu}) for any given
%AP5: geometry of NS rotation, NS mass, EoS and emitted neutrino energy.
rotation geometry, mass, EoS of the star and emitted neutrino energy.

To obtain the theoretical distribution of neutrino pulsars over the amplification factors $f(a_\nu)$ we perform Monte Carlo simulations. 
In each simulation, we construct a neutrino pulse profile and calculate apparent neutrino luminosity for NS inclination 
\beq 
i = \mathrm{arccos}(1-2X_1)
\eeq
and magnetic obliquity 
\beq 
\theta_B = \pi X_2,
\eeq
where $X_1,\,X_2\in(0;1)$ are random numbers.
The constructed
%AP5: distribution function is normalized as $\int_0^{\infty}f(a_\nu)\d a_\nu=1$. In practice it is useful to consider the
differential  distribution function $f(a_\nu)$ is normalized as
\beq
\int_0^{\infty}f(a_\nu)\d a_\nu=1.
\eeq
 In practice it is useful to consider the cumulative distribution
function describing the fraction of objects amplified by a factor larger than $a_\nu$:
\beq\label{eq:F(a)}
{\cal F}(a_\nu)\equiv \int\limits_{a_\nu}^{\infty} f(x)\d x. 
\eeq 

%%%%%%%%%%%%%%%%%%%%%%%%%%%%%%%%%%%%%%%%
\section{Numerical results}
\label{sec:NumRes}
%%%%%%%%%%%%%%%%%%%%%%%%%%%%%%%%%%%%%%%%

In this section we demonstrate results of our numerical simulations of neutrino trajectories (Section \ref{sec:ResNeutrinoTraj}), angular distribution of neutrino energy flux
%AP5: [I removed "see" from the brackets]
(Section \ref{sec:ResAngDistrib}) and theoretical distributions of neutrino pulsars over the neutrino amplification factor (Section \ref{res:ResAmpDistrib}). 
The gravitational bending of neutrinos propagating through a star is affected by the internal mass distribution. 
We analyse mass density distributions calculated for three specific NS EoSs (see Section \ref{sec:NS_EoS}), assuming NS masses of $1.4M_\odot$ and $2M_\odot$, and for a strange star, assuming its mass of $1.4M_\odot$.
%AP5: Furthermore, 
In the latter case, to demonstrate the possible impact of neutrino
%AP5: scattering or 
absorption in a compact star, we perform simulations for neutrinos of different energies.

%%%%%%%%%%%%%%%%%%%%%%%%%%%%%%%%%%%%%%%%
\subsection{Neutrino trajectories}
\label{sec:ResNeutrinoTraj}
%%%%%%%%%%%%%%%%%%%%%%%%%%%%%%%%%%%%%%%%

\begin{figure}
\centering 
    \includegraphics[width=9.cm]{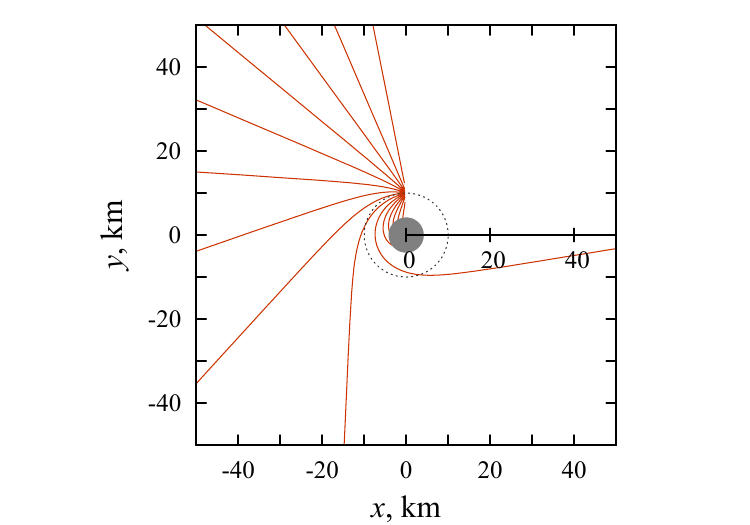} 
\caption{
Examples of neutrino trajectories calculated for particles emitted at the distance $10^6$~cm 
from the Schwarzschild black hole with mass $M=1.4\,M_\odot$.
%AP5: Dark grey region illustrates the
The grey circular area is the region
surrounded by the Schwarzschild radius $R_\mathrm{Sch}$, while the
%AP5: region bounded by the  blue dot circle shows the volume within radius of $10^6\,{\rm cm}$.
blue dotted circle marks the distance $r=10$ km from the centre.
}
\label{pic:sc_nu_tr_BH}
\end{figure}

\begin{figure}
\centering 
\includegraphics[width=9.cm]{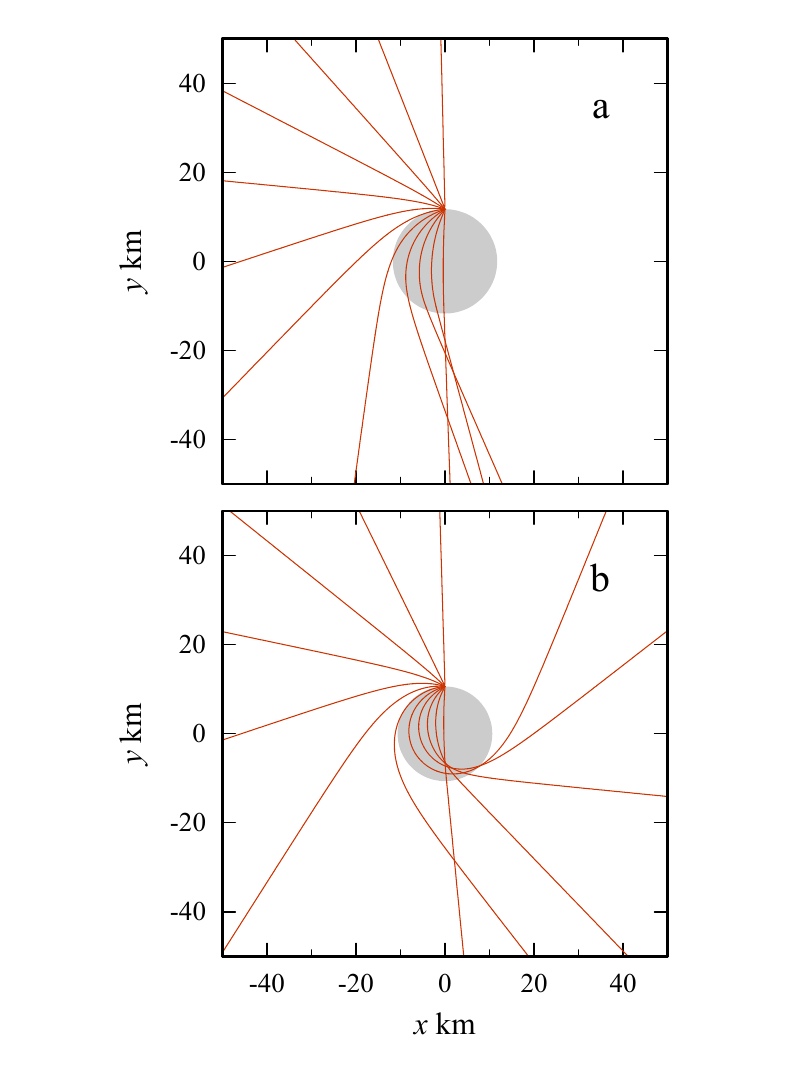}
\caption{
Examples of neutrino trajectories calculated for particles emitted at the NS surface for the case of different NS mass (and radius):  
(a) $M=1.4M_\odot$,  
(b) $M=2M_\odot$.  
In these calculations, the SLy4 EoS is assumed. For comparison, the same uneven steps for the initial co-latitudes of trajectories are chosen in both panels.
}
\label{pic:sc_nu_tr_NS}
\end{figure}

\begin{figure}
\centering 
\includegraphics[width=9.cm]{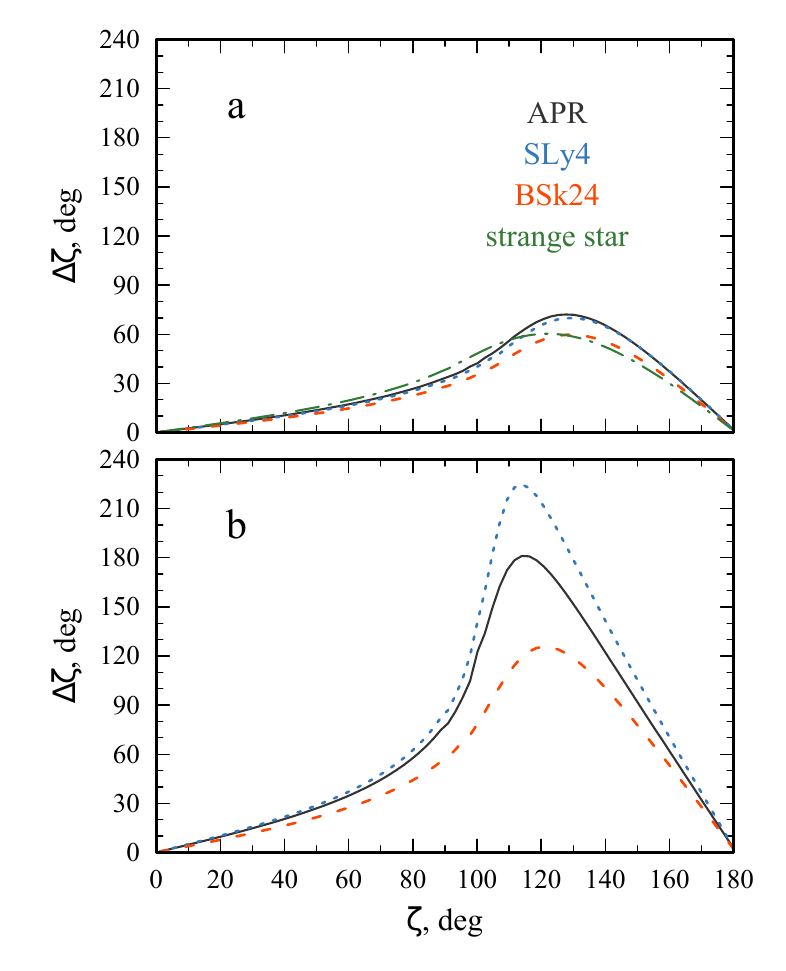}
\caption{
The deflection angle for neutrinos emitted from the surface of a NS at different directions given by angle $\zeta$ (see Section\,\ref{sec:traj}).
Different curves are calculated for different EoSs: 
APR (solid black),
SLy4 (dotted blue),
BSk24 (dashed red),
and for the case of strange star (dashed-dotted green).
The upper and lower panels are given for NS mass $1.4M_\odot$ and $2M_\odot$ respectively.
}
\label{pic:sc_NS_nu_def_ang_}
\end{figure}

Examples of neutrino trajectories calculated in the Schwarzschild metric near a $1.4M_\odot$ black hole are shown in Fig.\,\ref{pic:sc_nu_tr_BH}.
Fig.\,\ref{pic:sc_nu_tr_NS} depicts neutrino trajectories emitted from the surface of $1.4M_\odot$ (upper panel) and $2M_\odot$ (lower panel) NS.
Unlike photons, neutrinos can penetrate into
%AP: an NS, 
a compact star,
where their trajectories follow geodesic paths.
Within
%AP: a compact object, 
the star,
neutrino trajectories are influenced by the mass distribution,
%AP5: internal 
gravitational potential, and pressure (see Appendix \ref{app:GR_DU} for details).
The greater the mass of a star and the more concentrated its matter is toward the center, 
the larger the deviation of particles from their original propagation direction, 
i.e. the deflection angle (see Fig.\,\ref{pic:sc_NS_nu_def_ang_}).

%-----------------------------------------%
%-----------------------------------------%

%-----------------------------------------%
%-----------------------------------------%

%-----------------------------------------%
%-----------------------------------------%

%%%%%%%%%%%%%%%%%%%%%%%%%%%%%%%%%%%%%%%%%%%%%%%%%%
\subsection{Angular distribution of neutrino flux}
\label{sec:ResAngDistrib}
%%%%%%%%%%%%%%%%%%%%%%%%%%%%%%%%%%%%%%%%%%%%%%%%%%

\begin{figure}
\centering 
\includegraphics[width=8.5cm]{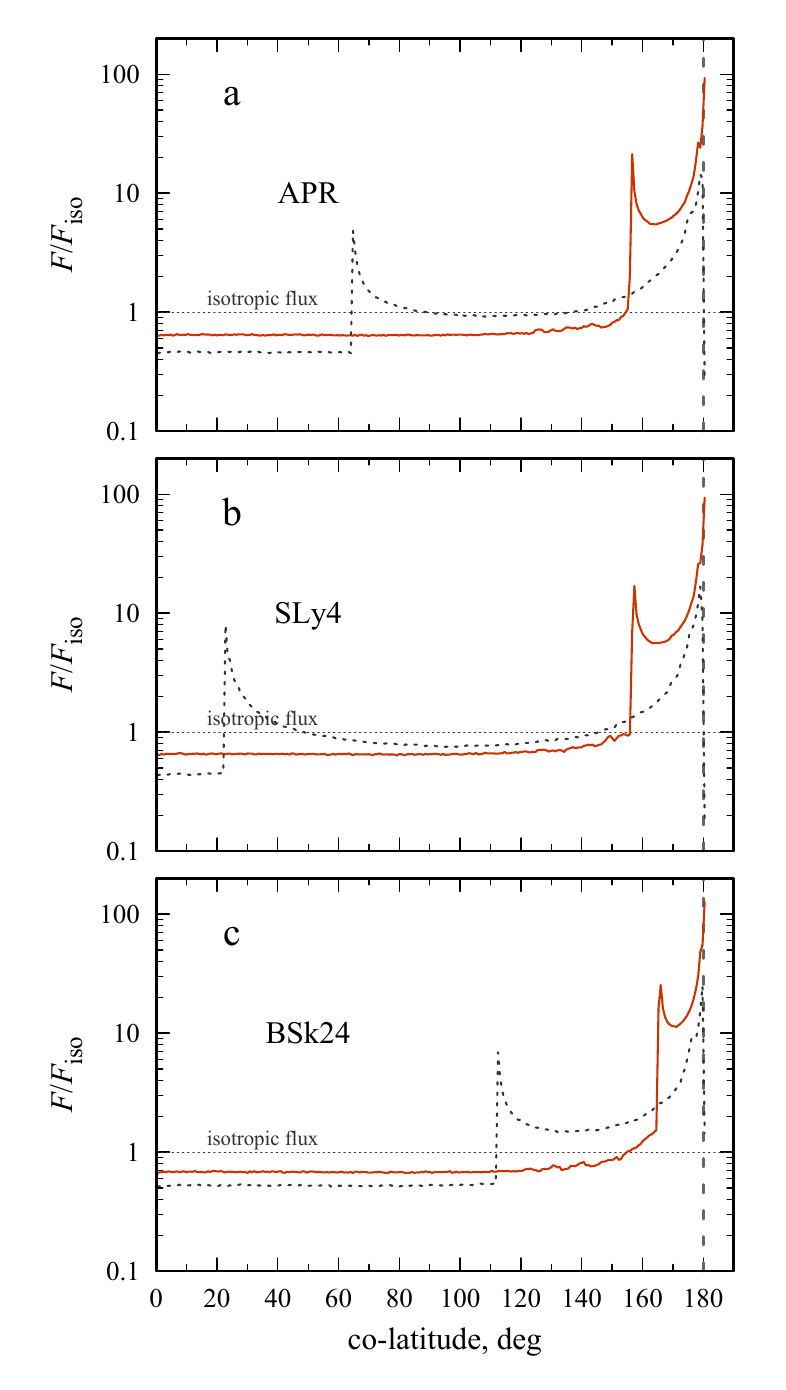} 
\caption{
Angular distribution of neutrino energy flux from one of the poles of a NS is given by red solid (black dotted) line for $1.4M_\odot$ ($2M_\odot$) NS.
The horizontal dotted line shows the level of the isotropic neutrino energy flux. 
Different panels correspond to different equations of state: 
(a) ARP, 
(b) SLy4,
(c) BSk24.
In the case of relatively cold NSs under consideration, neutrino energy does not affect neutrino transfer within a star and, thus, the angular distribution.
}
\label{pic:sc_NS_nu_flux_real}
\end{figure}

Utilizing the calculated neutrino trajectories, we derive the angular distribution of neutrino energy flux in the reference frame of a NS (i.e., in a frame where the star does not rotate).
Neutrino trajectories started from the magnetic pole at the NS surface are curved and tend to converge in certain directions, leading to a significant amplification
%AP5: (by a factor of more than 10)
of the neutrino energy flux in those areas. 
The directions of enhanced flux depend on the NS mass and internal structure, which are governed by the EoS.
We note that the angular distributions of neutrino flux always show two peaks.
The first
%AP5:
peak is
in the direction opposite to the magnetic pole of a star that produces neutrinos, i.e. at co-latitude $\sim \pi$ (see Fig.\,\ref{pic:sc_NS_nu_flux_real}).
Similar peaks have been reported earlier for photons lensed in the gravitational field of a NS (see Fig.\,3,\,4 and 9 in \citealt{1988ApJ...325..207R},
Fig.\,8,\,9 in \citealt{2001ApJ...563..289K},
Fig.\,10--12 in \citealt{2018MNRAS.474.5425M}, and
Fig.\,6 in \citealt{2024MNRAS.530.3051M}).
The second peak in the angular distribution
%AP5: of neutrinos 
corresponds to neutrinos experiencing the maximal deflection $\Delta\zeta$ (see Fig.\,\ref{pic:sc_NS_nu_def_ang_}), 
at co-latitude
%$\sim (\zeta+\Delta\zeta)$ (see Fig.\,\ref{pic:sc_NS_nu_flux_real}).
$\sim [2\pi - (\zeta+\Delta\zeta)]$ (see Fig.\,\ref{pic:sc_NS_nu_flux_real}).
The angular distribution of neutrinos depends on both the EoS (compare different panels in Fig.\,\ref{pic:sc_NS_nu_flux_real}) and the mass of a NS (compare solid red and dotted black lines in Fig.\,\ref{pic:sc_NS_nu_flux_real}). 
For NSs with smaller masses, the maximum enhancement of the neutrino flux is more pronounced. 
In contrast, more massive NSs deflect neutrinos more strongly from their original propagation direction. 
It results in the angular distribution that is closer to the isotropic one, albeit with distinct peaks still present.

%-----------------------------------------%
%-----------------------------------------%

\begin{figure}
\centering 
\includegraphics[width=9.cm]{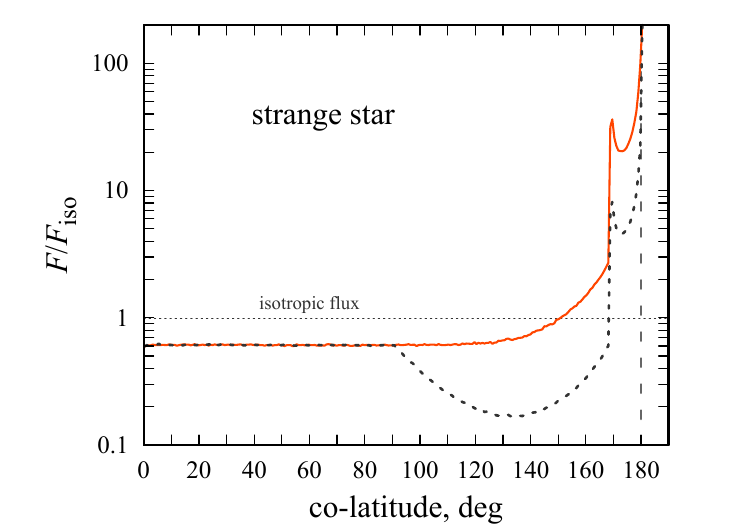} 
\caption{
The angular distribution of neutrino energy flux from one of the poles of a strange star
with $M=1.4M_\odot$ and $M_r$ conforming with the  EoS proposed by \citet{2000A&A...359..311Z}.
Different lines are calculated for different neutrino energy: $100$~keV (solid red) and 1~MeV (dotted black).
}
\label{pic:sc_NS_nu_flux_real_QS}
\end{figure}

In the case of strange stars, unlike the NSs, neutrino absorption can
be noticeable. 
Nevertheless, angular distribution becomes strongly anisotropic and neutrino energy flux can exceed the isotropic flux by more than an order of magnitude (see Fig.\,\ref{pic:sc_NS_nu_flux_real_QS}).
Only at high energies $\sim 1$~MeV, some fraction of neutrinos is absorbed, which reduces the flux  
%in the direction opposite to the location of emitting pole of a star 
directed towards a star (compare solid and dotted lines in Fig.\,\ref{pic:sc_NS_nu_flux_real_QS}).

%-----------------------------------------%
%-----------------------------------------%

%%%%%%%%%%%%%%%%%%%%%%%%%%%%%%%%%%%%%
\subsection{Luminosity function}
\label{res:ResAmpDistrib}
%%%%%%%%%%%%%%%%%%%%%%%%%%%%%%%%%%%%%

\begin{figure}
\centering 
\includegraphics[width=8.3cm]{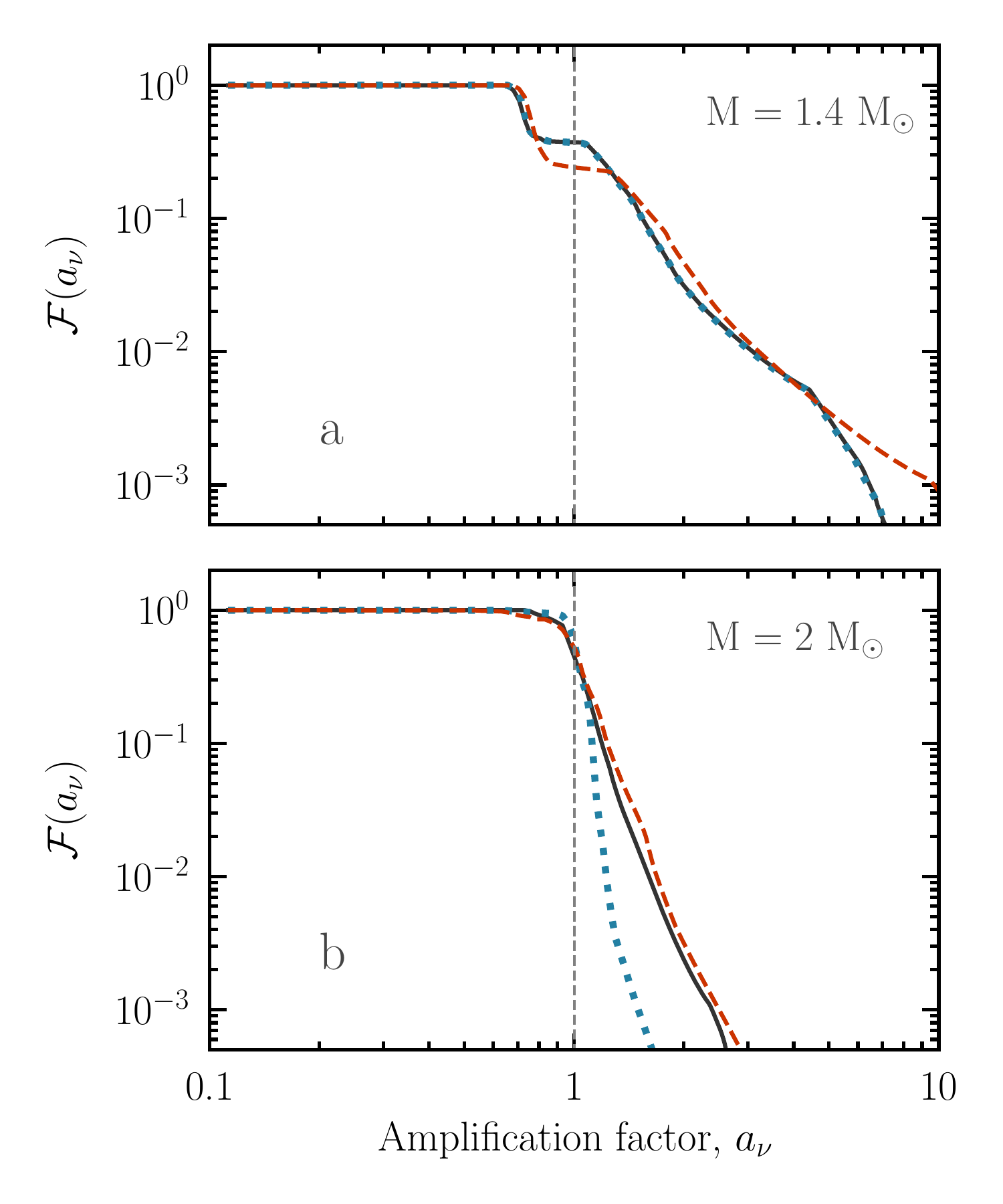} 
\caption{
Fraction of neutrino pulsars of amplification factor exceeding $a_\nu$.
Different lines show results calculated for  different EoSs:
APR (solid black), SLy4 (dotted blue) and BSk24 (dashed red).
Different panels correspond to different masses of a NS: 
(a) $M=1.4\,M_\odot$, 
(b) $M=2\,M_\odot$.
One can see that large mass of a NS reduces significantly a fraction of strongly amplified sources. 
}
\label{pic:sc_NS_nu_dist_real}
\end{figure}

Using the calculated angular distributions of neutrino energy flux, we derive theoretical distributions of NSs over the neutrino amplification factor $a_\nu$ and calculate the fraction ${\cal F}(a_\nu)$ of NSs with amplification factors above specific values according to equation~(\ref{eq:F(a)}), as  described in subsection~\ref{sec:Num_amp_fac}. 
These distributions are shown in Fig.\,\ref{pic:sc_NS_nu_dist_real}.

One can see that the distributions of neutrino pulsars over the amplification factor are relatively restricted: the majority of objects exhibit amplification factors within the interval $a_\nu\in(0.5,10)$.
The anticipated population of objects with relatively large amplification factors decreases for larger NS masses (see Fig.\,\ref{pic:sc_NS_nu_dist_real}).
For the considered EoSs, only 
%$\sim 0.05\%$ ($\sim 0.1\%$) 
$\sim 0.1\% $\  ($\sim 0.05\%$) of neutrino pulsars demonstrate an amplification factor $a_\nu>10$ for NS masses of $1.4M_\odot$ ($2M_\odot$).
The expected distribution of objects over the amplification factor depends on the EoS insignificantly  
(compare different lines in Fig.\,\ref{pic:sc_NS_nu_dist_real}).

%-----------------------------------------%
%-----------------------------------------%

\begin{figure}
\centering 
\includegraphics[width=8.7cm]{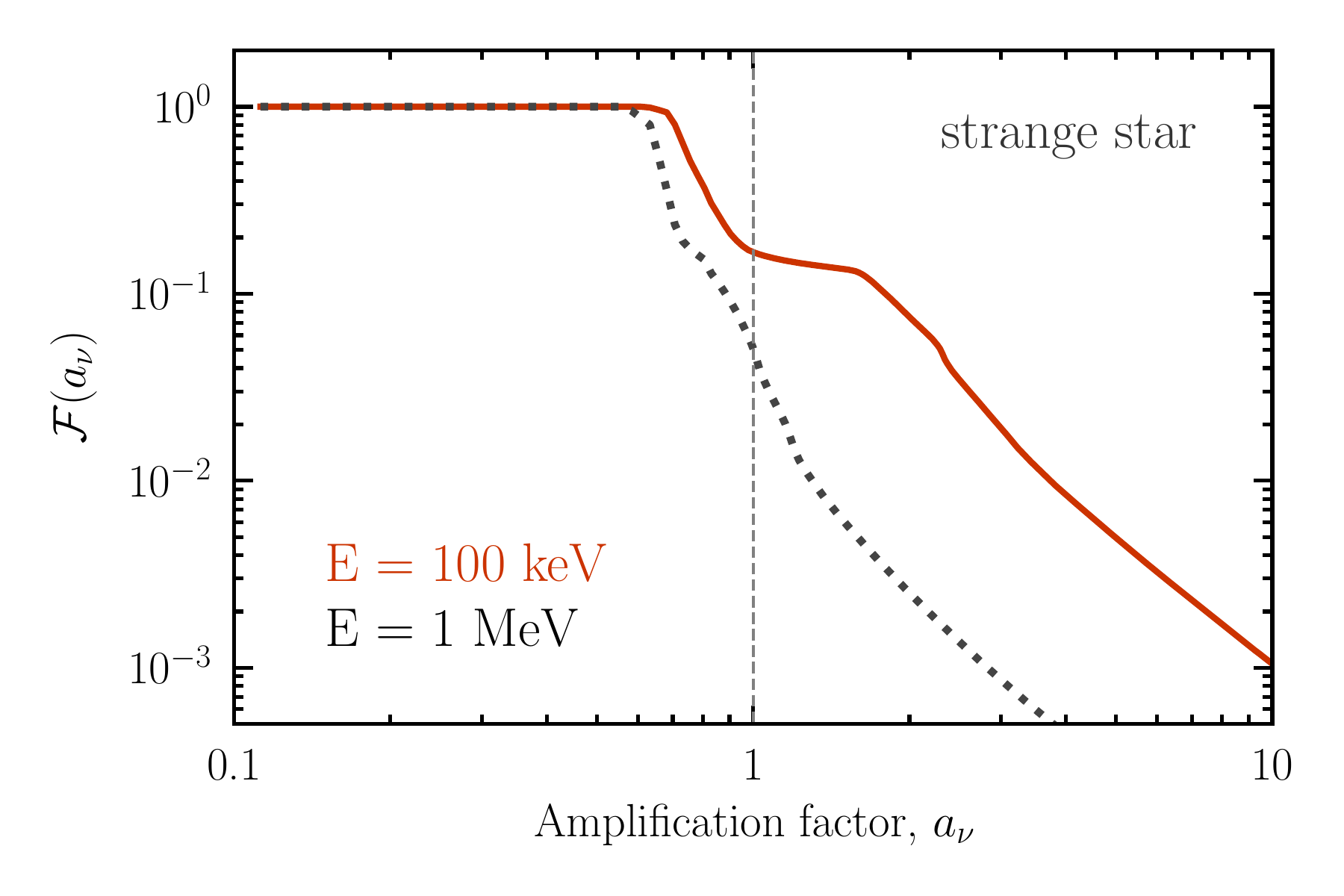} 
\caption{
Fraction of neutrino pulsars of amplification factor exceeding $a_\nu$ calculated for the case of a strange star with $M=1.4M_\odot$ and $M_r$ conforming with the  EoS proposed by \citet{2000A&A...359..311Z}.
Different lines are calculated for different neutrino energy: 100~keV (solid red) and 1~MeV (dotted black).
The line corresponding to higher neutrino energy shows smaller fractions because of neutrino absorption in a star.
}
\label{pic:sc_fL_neutrino_E_QS}
\end{figure}

In the case of strange stars, the distribution of objects over the amplification factor depends on neutrino energy.
About $10\%$ of ULX hosting strange stars can demonstrate amplification factor $a_\nu>2$ and $\sim 0.01\%$ can show amplification factors $a_\nu>10$ (see Fig.\,\ref{pic:sc_fL_neutrino_E_QS}).

%-----------------------------------------%
%-----------------------------------------%

%%%%%%%%%%%%%%%%%%%%%%%%%%%%%%%%%%%%%%
\section{Summary}
\label{sec:Summary}
%%%%%%%%%%%%%%%%%%%%%%%%%%%%%%%%%%%%%%

We have explored the impact of gravitational bending on neutrino emission in strongly magnetized NSs undergoing extreme mass accretion rates, such as bright X-ray transients or ULX pulsars.
NS interiors in the considered class of objects are cold enough (temperature $\lesssim 10$~keV) to be completely transparent to neutrino emission in keV and MeV energy bands \citep{1987A&A...179..127H}.
Thus, a fraction of neutrino emission is going through a NS experiencing gravitational bending.
Through Monte Carlo simulations in the metric generated by spherically symmetric and quasi-static mass distribution within a NS, we simulated neutrino beam patterns (Figs.\,\ref{pic:sc_NS_nu_flux_real},\,\ref{pic:sc_NS_nu_flux_real_QS}) influenced by neutrino gravitational bending.
The gravitational bending induces strong anisotropy in neutrino emission within the NS reference frame, leading to the phenomenon of neutrino pulsars.

{Using calculated beam patterns, we have obtained the theoretical distributions of neutrino pulsars over the amplification factors (\ref{eq:a_nu}) that show the ratio of apparent (\ref{eq:L_ave}) and actual luminosity in neutrinos} (see Fig.\,\ref{pic:sc_NS_nu_dist_real} and \ref{pic:sc_fL_neutrino_E_QS}).
These distributions reveal limited ranges of amplification factors.
The majority of neutrino pulsars are expected to fall within the interval $a_\nu\in(0.5,10)$.
For the considered equations of state, only approximately 
%$\sim 0.05\%$ ($\sim 0.1\%$) 
$\sim 0.1\%$ ($\sim 0.05\%$)
of neutrino pulsars exhibit an amplification factor $a_\nu>10$ at a neutron star mass of $1.4M_\odot$ ($2M_\odot$).
Thus, the expected neutrino flux from known pulsating ULXs and bright Be X-ray transients most likely remain to be below the isotropic neutrino background even in the case of flux amplification due to neutrino gravitational bending (see previous estimations that neglect gravitational bending in \citealt{2023MNRAS.522.3405A}).

In the case of strange stars, where the core of a star is composed of quark matter, high energy neutrinos can be subject to absorption.
As a result, neutrino beam pattern becomes energy dependent (see Fig.\,\ref{pic:sc_NS_nu_flux_real_QS}), which affects the expected distribution of objects powered by accretion onto strange start over the amplification factor (see Fig.\,\ref{pic:sc_fL_neutrino_E_QS}).
Note that
presence of quark matter inside a star can cause
some additional
heating of the stellar interiors by the source of neutrino emission of the surface
due to the neutrino absorption in such matter.

%%%%%%%%%%%%%%%%%%%%%%%%%%%%%%
\section*{Acknowledgements}
%%%%%%%%%%%%%%%%%%%%%%%%%%%%%%

The authors thank Simon Portegies Zwart for discussions.
We are grateful to an anonymous referee for their useful comments and suggestions which helped us fix a mistake in the original version of manuscript and improve the paper.
AAM thanks UKRI Stephen Hawking fellowship.
The work of AYP and IDM was partially supported by the Ministry of Science and Higher Education of the Russian Federation (Agreement No.\,075-15-2024-647).

%%%%%%%%%%%%%%%%%%%%%%%%%%%%%%%%%%%%%
\section*{Data availability}
%%%%%%%%%%%%%%%%%%%%%%%%%%%%%%%%%%%%%

The calculations presented in this paper were performed using a private code developed and owned by the corresponding author. All the data appearing in the figures are available upon request.

%%%%%%%%%%%%%%%%%%%%%%%%%%%%%%%%%%%%%%%%%%%%%%%%%%%%%%%%%%%%%%%%%%%%%%%%%%%%%%
%% Bibliography %%
%%%%%%%%%%%%%%%%%%%%%%%%%%%%%%%%%%%%%%%%%%%%%%%%%%%%%%%%%%%%%%%%%%%%%%%%%%%%%%
%\bibliographystyle{mn2e}
%\bibliographystyle{mnras}
%\bibliography{allbib}

%%%%%%%%%%%%%%%%%%%%%%%%%%%%%%%%%%%%%%%%%%%%%%%%%
\appendix
%%%%%%%%%%%%%%%%%%%%%%%%%%%%%%%%%%%%%%%%%%%%%%%%%

%%%%%%%%%%%%%%%%%%%%%%%%%%%%%%%%%%%%%%%
\section{Geodesic lines}
\label{app:GR_DU}
%%%%%%%%%%%%%%%%%%%%%%%%%%%%%%%%%%%%%%%

Equation (\ref{eq:DE_traj}) can be rewritten in a form appropriate for numerical integration
as follows:
\beq  
  \frac{\d r}{\d\varphi} = \pm\,r^2 \left( \frac{1}{A(r)B(r)\,b^2} - \frac{1}{A(r)\,r^2} \right)^{\!1/2}.
\label{eq:DE_traj1_}
\eeq
The numerical modelling of neutrino trajectories based on this first-order differential equation requires a correct choice of the sign on the right-hand side, as described in Appendix~\ref{app:TrNum}. One can avoid this sign ambiguity by using the second-order equation
\begin{multline}
\label{eq:second_order_Null_geodesic_sph_symm}
    \frac{2A(r)}{r^4}\frac{\d^2 r}{\d\varphi^2}
    +\frac{1}{r^4} \left(\frac{\d A(r)}{\d r}-\frac{4A(r)}{r}\right)
    \left(\frac{\d r}{\d\varphi}\right)^{\!2}
    \\ -\frac{2}{r^3}+\frac{{\d B(r)}/{\d r}}{b^2B^2(r)}=0,
\end{multline}
which is obtained by taking the derivative of both sides of equation~(\ref{eq:DE_traj}) over $r$.

Based on the Appendix in \citet{2002ApJ...566L..85B}, we can express $b$ in terms of the trajectory variables. 
The tangent vector for null geodesic line associated with the trajectory in the metric (\ref{eq:Sph_sym_metr}) 
can be written as
\beq\label{eq:null_geo_app}
  u^\mu=\frac{\d x^\mu}{\d\lambda},
\eeq
where $\mu$ is the index of the coordinate $(t,r,\theta,\varphi)$ and $\lambda$ is an affine parameter. We can put $\theta=\pi/2$ without loss of generality. Killing vectors $\partial/\partial t$ and $\partial/\partial\varphi$ for (\ref{eq:Sph_sym_metr}) correspond to the integrals of motion $u_t$ and $u_\varphi=b$, respectively. If we put $\lambda=1/B(r)$, we get $u_t=-1$. Then from the condition $u^\mu u_\mu=0$ we obtain
\beq\label{eq:u_r}
  (u^r)^2=\frac{1}{A(r)B(r)}-\frac{b^2}{A(r)\,r^2}.
\eeq

Let us consider the massless particle at the radius $r$ and denote the angle between the particle momentum and the radial vector from the center of symmetry as $\zeta$. Then
\beq\label{eq:tg_a}
\tan \zeta =\left(\frac{u^\varphi u_\varphi}{u^r u_r}\right)^{\!1/2}=\frac{b}{r}\left(\frac{1}{B(r)}-\frac{b^2}{r^2}\right)^{-1/2}.
\eeq
Therefore, $b$ can be related to $r$ and $\zeta$ as follows:
\beq\label{eq:b}
  \sin\zeta  =\frac{b}{r}\sqrt{B(r)}.
\eeq
Since the vector fields $\partial/\partial t$ and $\partial/\partial \varphi$ are the Killing fields both for the Schwarzschild metric and (\ref{eq:Sph_sym_metr}), the value of $b$ does not change if the particle crosses the neutron star surface. Thus we arrive at equation~(\ref{eq:beta2zeta}).

Let us consider opacity transformation in General Relativity. Neutrino transport process can be described by the relativistic Boltzmann equation for massless particles that can be written as \citep{1966AnPhy..37..487L}:
\beq\label{eq:GR_RT}
  k^\alpha\frac{\partial \mathcal{I}}{\partial x^\alpha}-\Gamma^\alpha_{\beta\gamma}k^\beta k^\gamma\frac{\partial \mathcal{I}}{\partial k^\alpha}=\mathcal{J}-\kappa \mathcal{I}. 
\eeq
Here, $\Gamma^\alpha_{\beta\gamma}$ are the Christoffel symbols, $k^\alpha$ is the particle four momentum, and $\mathcal{I}=I_\nu(\mathbf{\Omega})/\nu^3$ is the invariant specific intensity. 
The ordinary specific intensity $I_\nu(\mathbf{\Omega})$ is usually defined in relation to the radiative transfer \citep[e.g.,][]{1984frh..book.....M}, where $\nu$ is the photon frequency and $\mathbf{\Omega}$ is the photon propagation direction. 
Note that in General Relativity the photon frequency can be defined only in the local rest frame associated with an observer. 
In our case, the frequency is $\nu=k^{\hat{0}}/h$, where $h$ is the Planck constant and $k^{\hat{0}}$ is the neutrino energy measured in the reference frame of the observer whose $(r,\theta,\varphi)$ coordinates do not change. 
Since $\partial/\partial t$ is the Killing vector for the spherical static metric (\ref{eq:Sph_sym_metr}), $\nu\sqrt{B(r)}=\mathrm{constant}$. 

Furthermore, $\mathcal{J}=j_\nu/\nu^2$ in equation~(\ref{eq:GR_RT}) is an invariant emissivity,
$j_\nu$ being an ordinary emissivity;
$\kappa=\nu\alpha_\nu$ is an invariant absorption coefficient, $\alpha_\nu\propto 1/\lambda_\nu$,
being an ordinary absorption coefficient, and $\lambda_\nu$ is a mean free path at the frequency $\nu$. The quantities $j_\nu$, $\alpha_\nu$ and $\lambda_\nu$ are defined in the same reference frame as the frequency $\nu$. 
For an accurate calculation of neutrino transfer in neutron stars it is necessary to take into account the transformation of mean free path along the geodesic line due to the change of the metric coefficient $B(r)$. 
Since the typical mean free path
of a neutrino with energy of a few hundred keV 
is very large in comparison with the typical 
NS radius (see Section~\ref{sec:opac})
we can neglect neutrino opacities in the  NSs. In the quark stars, we can neglect neutrino scattering, but should take into account neutrino absorption.

In our numerical model, we trace the motion of each individual neutrino as it propagates through an NS.  
Let us consider a neutrino moving from $(t, r, \pi/2, \varphi)$ to $(t + \d t, r + \d r, \pi/2, \varphi + \d\varphi)$ in Schwarzschild coordinates $(t, r, \theta, \varphi)$ (without loss of generality, we assume $\theta = \pi/2 = \mathrm{const}$).  
The spatial displacement vector $\bm{\delta l} = (\d r, 0, \d\varphi)$ lies in the tangent space at the point $(t, r, \pi/2, \varphi)$ and is represented in the coordinate basis as:  
\beq
\bm{\delta l} = \d r \, \frac{\partial}{\partial r} + \d\varphi \, \frac{\partial}{\partial\varphi}.
\eeq   
The optical depth $\d\tau$ associated with this infinitesimal displacement $\bm{\delta l}$ is $\d\tau = \d\overline{s}/\lambda$, where $\lambda$ is the neutrino mean free path and $\d\overline{s}$ is the length of the spatial motion. Both quantities are evaluated in the local Minkowski frame corresponding to the element of matter with which the neutrino interacts.

We assume the neutron star matter is at rest; therefore, the local orthonormal basis is:  
\begin{multline}
   \mathbf{e_{(0)}} = \frac{1}{\sqrt{B(r)}}\frac{\partial}{\partial t},
\quad
   \mathbf{e_{(1)}} = \frac{1}{\sqrt{A(r)}}\frac{\partial}{\partial r},
\\
   \mathbf{e_{(2)}} = \frac{1}{\sqrt{r}}\frac{\partial}{\partial \theta},
\quad
   \mathbf{e_{(3)}} = \frac{1}{\sqrt{r\sin^2(\theta)}}\frac{\partial}{\partial \varphi}.
\end{multline}  
In this basis, the displacement vector is:  
\[
\bm{\delta l} = \sqrt{A(r)}\d r \, \mathbf{e_{(1)}} + r \d\varphi \, \mathbf{e_{(3)}}.
\]  
Its length is:  
\beq
  \d\overline{s} = \sqrt{A(r)(\d r)^2 + r^2(\d\varphi)^2}.
\label{eq:ds}
\eeq 

Note that $\d\overline{s}$ is numerically identical to the length computed in the spatial part of the spherically symmetric
%AP5: metric:  
%AP5: \[
%AP5: \d s^2 = -B(r)\d t^2 + A(r)\d r^2 + r^2(\d\theta^2 + \sin^2\theta \d\varphi^2),  
%AP5: \]  
metric (\ref{eq:Sph_sym_metr}),
because $\mathbf{e_{(0)}}$ is parallel to $\partial/\partial t$, as the neutron star matter is at rest.  

The mean free path depends on the neutrino energy, which changes along the geodesic due to gravitational redshift: $E_\nu \propto 1/\sqrt{B(r)}$. This effect is fully accounted for in our Monte Carlo modeling.

%%%%%%%%%%%%%%%%%%%%%%%%%%%%%%%%%%%%%%%
\section{{Simulations of neutrino trajectories}}
\label{app:TrNum}
%%%%%%%%%%%%%%%%%%%%%%%%%%%%%%%%%%%%%%%

We calculate neutrino trajectories, described by differential equation (\ref{eq:DE_traj}). were the mass distribution is spherically symmetric and given by $M_r$.
The impact factor can be calculated from the initial direction of particle motion according to equation (\ref{eq:beta2zeta}).
A trajectory is determined by the
initial coordinates of a particle 
$\bm{r}_0$
and initial direction of its motion, which is given by the unit vector of particle velocity 
\beq\label{eq:e_v0}
\bm{e}_{v0}=\frac{\bm{v}_0}{|\bm{v}_0|}.
\eeq
Simulating a trajectory, we choose a spacial separation between the nearest two points of approximate trajectory $\Delta s$ and follow the steps:\\
\begin{enumerate}[leftmargin=12pt]
\item
Using the starting point of particle trajectory $\bm{r}_0$ and the direction of its initial velocity given by the unit vector $\bm{e}_{v0}$ (\ref{eq:e_v0}), we calculate the second point of approximate trajectory:
\beq
\bm{r}_1 = \bm{r}_0 + \bm{e}_{v0}\Delta s.
\eeq 
At this step $i=1$.
\item 
Then we get the angle between positions $\bm{r}_i$ and $\bm{r}_{i-1}$:
\beq 
\cos\Delta\varphi_{i,i-1} = \frac{(\bm{r}_i,\bm{r}_{i-1})}{|\bm{r}_i|\,|\bm{r}_{i-1}|},
\eeq 
where $(\bm{r}_i,\bm{r}_{i-1})=\sum_{j=1}^3 r^{(j)}_i r^{(j)}_{i-1}$ denotes the scalar productions of two vectors and $r_i^{(j)}$ is $j^\mathrm{th}$ Cartesian coordinate of vector $\bm{r}_i$.
\item\label{step:3rd_point}
We get direction towards the $(i+1)^\mathrm{th}$ point of approximate particle trajectory:
\beq 
\bm{e}_{r,i+1} = 
\frac
{\bm{r}_i + \bm{e}_{v,i-1}\Delta s}
{|\bm{r}_i + \bm{e}_{v,i-1}\Delta s|}
\eeq 
and the angle between $\bm{e}_{r,i+1}$ and $\bm{r}_i$:
\beq
\Delta\varphi_{i+1,i}^*=\frac{(\bm{e}_{r,i+1},\bm{r}_i)}{|\bm{r}_i|}.
\eeq
\item 
Using the second-order Runge-Kutta method, applied to the differential equation (\ref{eq:DE_traj1_}),
we compute the radial distance $r_{i+1}$ at the next step of the simulation.
The sign on the right-hand side of (\ref{eq:DE_traj1_}) is determined based on whether the particle is moving toward or away from the center of the star.
If the right-hand side of (\ref{eq:DE_traj1_}) becomes zero at any step, the sign changes in the next step. 
This corresponds to the particle reaching its minimum distance from the center for a given impact parameter $b$.
\item
We get an estimation of the radial distance towards a new point of particle trajectory $r_{i+1}\simeq 0.5 R_\mathrm{Sch}(M_i)/u_{i+1}$ and calculate its position: 
\beq 
\bm{r}^*_{i+1}=r_{i+1}\bm{e}_{r,i+1}.
\eeq 
\item 
Because we want to get trajectory approximated by segments of a fixed length $\Delta s$, we recalculate the position of the latest point of neutrino trajectory as
\beq 
\bm{r}_{i+1} = \bm{r}_i 
+
\frac{\bm{r}^*_{i+1} - \bm{r}_{i}}{| \bm{r}^*_{i+1} - \bm{r}_{i}|} \Delta s.
\eeq 
The unit vector of neutrino velocity at the latest segment of trajectory is given by
\beq\label{eq:e_v_fin}
\bm{e}_{v,i}=
\frac{ \bm{r}_{i+1} - \bm{r}_i }
{|\bm{r}_{i+1} - \bm{r}_i|}.
\eeq 
\item 
We stop trajectory simulation if the particle experiences scattering at a given coordinate or if it is far from the central compact object: 
$|\bm{r}_{i+1}|>5\times 10^2R_\mathrm{Sch}$.
In this case, we have a final direction of particle motion given by (\ref{eq:e_v_fin}).
Otherwise, we return to step \ref{step:3rd_point} and continue the simulation of the trajectory.
\end{enumerate}
To control the accuracy of trajectory calculations, we perform it for smaller spacial step $\Delta s_1 = 0.5\,\Delta s$.
In the case of similar results of the simulation, we stop the improvement of accuracy.

The results of the performed algorithm outside a NS were verified by comparison of its results with the results of algorithms applied earlier
%AP5: in \citealt{2018MNRAS.474.5425M,2024MNRAS.530.3051M} 
by \citep{2018MNRAS.474.5425M,2024MNRAS.530.3051M}
to trace photon trajectories in X-ray pulsars. 

% Don't change these lines
\bsp % typesetting comment
\label{lastpage}
\end{document}